\documentclass[preprint,prd,nofootinbib,tightenlines,amsmath,amssymb,showpacs]{revtex4}
\usepackage{graphicx}
\usepackage{dcolumn}
\usepackage{bm}
\usepackage{color}
\usepackage{amsfonts}
\begin{document}
\baselineskip=15pt \parskip=3pt

\vspace*{3em}

\title{A covariantly foliated higher dimensional space-time: Implications for short distance gravity and BSM physics}

\author{Cao H. Nam}
\email{chnam@iop.vast.ac.vn} \affiliation{Institute of Physics,
Vietnam Academy of Science and Technology,\\  10 Dao Tan, Ba Dinh,
Hanoi, Vietnam}
\date{\today}

\begin{abstract}
We consider the space-time at short distances in which it is
described by a $D$-dimensional manifold (bulk) carrying out the
principal bundle structure. As a result, this space-time manifold
is foliated in the covariant way by the $(D-4)$-dimensional
submanifolds, realized as the space-like internal spaces, that are
smooth copies of the Lie group $G$ considered in this paper as the
special unitary group. The submanifolds being transversal to the
internal spaces are realized as the external spaces and in fact
identified as the usual $4$-dimensional world. The fundamental
degrees of freedom determining the geometrical dynamics of the
bulk corresponding with short distance gravity are given by the
gauge fields, the external metric field and the modulus fields
setting dynamically the volume of the internal spaces. These gauge
fields laying the bulk is to point precisely out the local
directions of the external spaces which depend on the topological
non-triviality of the space-time principal bundle. The physical
size of the internal spaces is fixed dynamically by the moduli
stabilization potential which completely arise from the intrinsic
geometry of the bulk. A detail description of the low energy bulk
gravity in the weak field limit is given around the classical
ground state of the bulk. Additionally, we investigate the
dynamics of the fundamentally $4$-dimensional Weyl spinor fields
and the fields of carrying out the non-trivial representations of
the Lie group $G$ propagating in the bulk in a detail study. These
results suggest naturally the possible solutions to some the
experimental problems of Standard Model, the smallness of the
observed neutrino masses and a dark matter candidate.
\end{abstract}

\pacs{04.50.-h, 02.40.-k, 04.60.-m, 11.25.Mj}

\maketitle

\section{\label{intro}Introduction and Proposal}
Our present understanding of the observable space-time
experimentally confirmed being valid to distances of the order
$10^{-16}$ cm is well provided by General Relativity (GR) of the
gravity. However, the gravitational phenomena only play the
important roles in the macroscopic world whereas in the quantum
world of the elementary particles they are not very significant
due to the gravitational coupling constant which is so small
compared those in Standard Model (SM). As well-known, all
nongravitational interactions are described by theory of
connections (or gauge fields) which are accommodated into the
quantum framework. This situation is completely different to that
for the gravitational interaction. Since it is possible that
unknown high energy gravitational effects should be contained in
the mysteriously deeper structures of the space-time. It should be
emphasized that the motivation for seeking of high energy
gravitational effects is also come from some other significant
reasons, such as physical phenomena associated with gravitational
collapse. These imply that the space-time at high energy regions
or short distances may in fact have dimensionality of more than
four as well as possess rather complex topologies and geometrical
structures.

The multidimensional universe possibility was first proposed by
Kaluza and Klein in attempting to unify the electromagnetic
interaction and the gravity in a $5$-dimensional space-time. The
extension tried to unify the forces of SM to the gravity can see
in \cite{Witten1981,Bailin-Love1987}. However, it is very
interesting in recent years that the physical models based on
extra-dimensions have devoted enormously to solve problems in
particle physics as well as cosmology
\cite{Antoniadis90,ADDmodel,RSImodel,UEDmodel}. The higher
dimensional extension of the space-time can thus offer a new way
of looking for Beyond the Standard Model (BSM) physics. It should
be noted that the dimensional reduction of the previously most
higher dimensional scenarios explicitly breaks the higher
dimensional covariance and the higher dimensional Poincar\'{e}
symmetry, for example, via the presence of $3-$brane\footnote{In
this case, $3-$brane is to correspond with the rigid object.
Consequently, the amplitude for a scattering process between two
particles which live on $3$-brane exchanged by gauge bosons
propagating into the bulk will be divergent even at tree level by
the contribution of the KK gauge boson modes. If, however,
$3$-brane is flexible, then braons describing the fluctuations of
$3$-brane in the extra dimensions will be included in the
computation of the amplitude. So contribution from the KK gauge
boson modes is automatically suppressed by an exponential factor
\cite{Bando1999}. In fact, the braons are to correspond with the
Goldstone bosons of translational invariance in extra dimensions
broken spontaneously
\cite{Kugo-Yoshioka2001,Dobado-Maroto2001,Alcaraz2003}. The
description of the effective theory for the flexible $3$-brane
universe is given in more detail by the work \cite{Sundrum1999}.
Some works also considered that this translational invariance is
possible an approximate symmetry, so some braons can have non-zero
masses. They always couple by pairs, since the lightest braon is
stable. It would also be difficult to detect them in observations.
Thus the braon can provide natural candidates to the dark matter
\cite{Cembranos2003}. Braon phenomenology at the LHC is given in
Ref. \cite{Cembranos-Delgado-Dobado2013}.} as
\begin{equation}
\mathbb{R}^{1,D-1}\times SO(D-1,1)\longrightarrow
\mathbb{R}^{1,3}\times SO(3,1)\times SO(D-4),
\end{equation}
where two first factors corresponds to translational symmetries
and the Lorentz group along the $3$-brane, and the second one does
to the rotation group along ($D-4$) extra dimensions which are
transverse to $3$-brane. Furthermore, a natural physical mechanism
to compact the extra dimensions and stabilize the corresponding
volume modulus fields is obviously still undetermined so far.

In this work, we consider a braneless higher dimensional
space-time $B^D$ of $D$-dimensions which is assumed a principal
bundle \cite{MDGfP,Nakahara}. It is in general factorized locally
as the topology product of two spaces, $U\times G$. Here $U$ is an
open subset of $4$-dimensional pseudo-Riemannian manifold $M^4$
with the Lorentz metric, and $G$ is a Lie group which we restrict
our attention to the special unitary group. Therefore, there
exists a diffeomorphism map $f$ smoothly deforming $U\times G$ to
a local neighbourhood $\pi^{-1}(U)\subset B^D$, which is
constructed as an inverse image of the surjective projection
\begin{equation}
\pi: B^D\longrightarrow M^4,\label{proj}
\end{equation}
as follows
\begin{equation}
f: U\times G\longrightarrow\pi^{-1}(U),\label{localtri}
\end{equation}
which is also called a local trivialization. Through such
diffeomorphic maps, it is quite clear to see that a set of points
belonging $B^D$ which corresponds with an inverse image
$\pi^{-1}(x)$ ($x\in M^4$), called the fibre at $x$, forms a
submanifold of $B^D$ that is a smooth copy of the Lie group $G$.
All such submanifolds are not necessary as $G$ and obviously
disjoint with together. Since the underline geometry of $B^D$ is
always given as the $D$-dimensional manifold foliated by
($D-4$)-dimensional submanifolds of which are smoothly equivalent
to the Lie group $G$. It is important that this foliation is
exactly to remain the higher dimensional covariance and
symmetries. These submanifolds will be realized as the space-like
internal spaces and also called the leaves of the foliation
\cite{Benjacu2006}. As indicated above, one can easily see that at
starting point we consider the internal spaces to be compact in
the sense of the mathematics. It is also important that
four-dimensional submanifolds whose vector fields point from one
fibre to another, thus being transversal to the internal spaces,
will be in fact identified as the usual four-dimensional world.
They are called the external spaces and in general locally smooth
equivalence to $M^4$. This means that the dynamics of the fields
in $B^D$ should correspond with the propagation along the internal
spaces and external spaces. However, in low energy limit we only
observe the motion from an internal space to another leading to
the dimensional reduction that is very natural to produce the
usual four-dimensional world as will be shown in this paper.
Effects of the dynamics along the internal spaces and others
coming from extra dimensions are elegantly hidden from the present
experiments on particle accelerators, astrophysics and cosmology.

Through the local description (\ref{localtri}), coordinates of a
point on $B^D$ are given as\footnote{Throughout this paper, the
indices $M,N,...$ running from $0$ to $D-1$ are used to denote the
higher dimensional indices, the $4$-dimensional external indices
are associated with $\mu,\nu,...=0,1,2,3$, and the internal
indices are labeled by $a,b,...=1,...,D-4$ or $i,j...=1,...,D-4$
as explained below.}
\begin{eqnarray}
X^M=x^\mu, \hspace*{0.5cm} M=0,1,2,3, \nonumber\\
X^M=\frac{\theta^a}{\Lambda}\equiv\hat{\theta}^a. \hspace*{0.5cm}
M=4,...,D-1.
\end{eqnarray}
Here $\{x^\mu\}\in \mathbb{R}^4$, and a set of real numbers
$\{\theta^a\}$ parameterizes each element of the Lie group G as,
$g=\textrm{exp}\{i\theta^aT_a\}$, with the generators $T_a$
($a=1,...,D-4$) satisfying the following commutation relation of
the Lie algebra $\mathfrak{g}$ of the Lie group $G$
\begin{equation}
[T_a,T_b]=if^c_{ab}T_c,
\end{equation}
where $f^c_{ab}$ are the structure constants. The coordinates
$x^\mu$ are realized as the $4$-dimensional external coordinates
while $\theta^a$ (or $\hat{\theta}^a$ ) are realized as the
internal coordinates, or fibre ones, which offer obviously the
global coordinate system on each internal space. It should be
noted that due to $\theta^a$ being dimensionless a new energy
scale $\Lambda$ characterizing physically to the internal spaces
is taken in the natural way. It should be noted that this scale is
not realized as the inverse radius of the compact internal spaces
which is in fact determined only if their metric is endowed. This
energy scale will be responsible for exciting the physical states
coming from the extra dimensions. Since it should be high enough.

The rule of the local coordinate transformations which corresponds
with a non-empty overlap of two any local neighbourhoods is
defined by
\begin{equation}
x'^\mu=\Lambda(x){^\mu}_\nu x^\nu,\
e^{i\theta'^aT_a}=h(x)e^{i\theta^aT_a},\label{gct}
\end{equation}
where $\Lambda(x)$ and $h(x)=\exp\{i\alpha^a(x)T_a\}$ are elements
of the linear transformation group $GL(4,\mathbb{R})$ and the Lie
group $G$, respectively.\footnote{$\theta'^a$ are explicitly
expressed in terms of $\theta^a$ and $\alpha^a(x)$ as
\begin{eqnarray}
\theta'&=&-i\ln\left(e^{i\alpha(x)}e^{i\theta}\right)
=-i\sum_{n=1}\frac{(-1)^{n-1}}{n}\left(e^{i\alpha(x)}e^{i\theta}-1\right)^n\nonumber\\
&=&\theta+\alpha(x)+\frac{i}{2}[\alpha(x),\theta]-\frac{1}{12}[\alpha(x),[\alpha(x),\theta]]+...,
\end{eqnarray}
where the internal coordinates and the internal transformation
parameters have been written in the compact form as,
$\beta=\beta^aT_a$, with $\beta$ referred to $\theta'$, $\theta$
and $\alpha(x)$.} Notice that, the rule of internal coordinate
transformation can be easily derived from (\ref{localtri}). As a
general principle, the invariant principle of the physical laws
requires that higher dimensional physical quantities have to be
covariant under the local coordinate transformations above.

This paper is organized as follows. In Sec. II, we present
essential settings of the proposed space-time. We study the
classical dynamics of pure bulk gravity in Sec. III. In Sec. IV,
the classical ground state of the bulk space-time is determined,
and the perturbative description of the bulk gravity in the low
energy limit is investigated around this background. In Sec. V, we
show how the $4$-dimensional Weyl spinor fields occur naturally on
$B^D$, and their dynamical Lagrangian is constructed in a
consistent way. As a result, we provide a understanding for the
natural smallness origin of the observed neutrino masses. In Sec.
VI, the dynamics of the fields of carrying out the local
non-trivial representations of the Lie group $G$ is considered.
Such a field is possible to suggest a dark matter candidate.
Finally, we devote to conclusions and comments in the last
section, Sec. VII.

\section{\label{Bgeo} Settings of Space-time $B^D$}
\subsection{\label{connone-form} The gauge fields of the space-time $B^D$}
Each local coordinate system on $B^D$ given above induces
naturally the basic vectors for the tangent spaces of $B^D$
presented by
\begin{equation}
\{\partial_M\}=\left(\{\partial_\mu\},\{\Lambda\partial_a\equiv\hat{\partial}_a\}\right),
\end{equation}
which rotate under the local coordinate transformations
(\ref{gct}) as follows
\begin{eqnarray}
\partial_\mu&\longrightarrow&\partial'_\mu=\frac{\partial x^\nu}{\partial x'^\mu}\partial_\nu+\frac{\partial\hat{\theta}^a}{\partial
x'^\mu}\hat{\partial}_a,\nonumber \\
\hat{\partial}_a&\longrightarrow&\hat{\partial}'_a=\frac{\partial\hat{\theta}^b}{\partial\hat{\theta}'^a}\hat{\partial}_b.\label{tagfrTrans}
\end{eqnarray}
With respect to $B^D$, each tangent space $T_XB^D$ at a given
point on it is always split into the following form
\begin{equation}
T_XB^D=H_XB^D\oplus V_XB^D,
\end{equation}
where $V_XB^D$ is called the vertical tangent subspace which is
tangent to an internal space whereas its complement, $H_XB^D$, is
called the horizontal tangent subspace which is tangent to an
external space. It is direct to see that vectors
$\hat{\partial}_a$ transform in the covariant way under the rule
of the fibre coordinate transformation. This means that the total
of $\{\hat{\partial}_a\}$ under the compatibility condition
corresponding to the second transformation of (\ref{tagfrTrans})
are equally good. Since they provide the well-defined local bases
for all vertical tangent spaces. On the other hand, the directions
along the internal spaces are defined independently of choosing a
local basis. However, the well-defined local bases for all
horizontal tangent spaces are not specified as seen more clearly
from the transformation in the first line of (\ref{tagfrTrans}).
This is as a result of that the directions to be transversal to
the fibres depend on the information of topological
non-triviality, or twisting, of the bundle $B^D$.

The principled way of how to move from an internal space to
another is given via the presence of a $\mathfrak{g}$-valued
one-form connection $\omega(X)$ which annihilates the horizontal
vectors. It is explicitly written in a local neighbourhood
$\pi^{-1}(U)$ as follows
\begin{equation}
\omega(X)=g^{-1}i\pi^{*}A(x)g+g^{-1}dg
 =ig^{-1}T_ag[d\theta^a+g_{\textrm{i}}A^a_\mu(x)dx^\mu],\label{conn}
\end{equation}
where $g=e^{i\theta^aT_a}$, $d$ is the exterior differential
operator, the pullback map $\pi^{*}$ is induced by the projection
operator $\pi$, $A^a_\mu$ with the dimension $[A^a_\mu]=1$ are the
gauge fields, and $g_{\textrm{i}}$ is a dimensionless constant (it
will play the role of the gauge coupling constant characterizing
the strength of the interaction mediated by the $A^a_\mu$).
Because of the pullback map $\pi^{*}$ acting on the local one-form
$A(x)=g_{\textrm{i}}A^a_\mu(x)T_adx^\mu$ lying on $M^4$, the gauge
fields $A^a_\mu$ in the last expression in Eq. (\ref{conn}) thus
live on $B^D$ in which they define topological non-triviality of
the space-time $B^D$ through their curvature tensor as indicated
below. The reader should not confuse $A^a_\mu$ in this framework
with the gauge fields in the usual gauge theories at which they
live on $M^4$ (in this case the space-time)
\cite{Daniel-Viallet1980}. Any curve,
$\overline{\gamma}(t)=(x^\mu(t),\hat{\theta}^a(t))$, flows
transversely to the fibres then each its tangent vector $V(t)$ is
horizontal one and thus is annihilated by the connection. Since we
can get the following relation
\begin{equation}
\frac{dg(t)}{dt}=-ig_{\textrm{i}}A^a_\mu
T_a\frac{dx^\mu(t)}{dt}g(t),\label{hlift1}
\end{equation}
where $g(t)=e^{i\theta^a(t)T_a}$. Then, we find
\begin{equation}
\frac{d\theta^a(t)}{dt}=-g_{\textrm{i}}A^a_\mu
\frac{dx^\mu(t)}{dt},\label{hlift2}
\end{equation}
that results in the formal expression for $V(t)$ as
\begin{eqnarray}
V(t)&=&\frac{dx^\mu(t)}{dt}\partial_\mu+\frac{d\hat{\theta}^a(t)}{dt}\hat{\partial}_a\nonumber\\
&=&\frac{dx^\mu(t)}{dt}(\partial_\mu-g_{\textrm{i}}A^a_\mu\partial_a)\equiv\frac{dx^\mu(t)}{dt}\hat{\partial}_\mu.
\end{eqnarray}
The total of $\{\hat{\partial}_\mu\}$ provide the well-defined
local bases for whole horizontal tangent spaces on $B^D$ if the
gauge fields $A^a_\mu$ transform as
\begin{equation}
A'_\mu(x')=\frac{\partial x^\nu}{\partial
x'^\mu}\left[h(x)A_\nu(x)h(x)^{-1}
+\frac{1}{ig_{\textrm{i}}}h(x)\partial_\nu
h(x)^{-1}\right],\label{gautrnI}
\end{equation}
where $h(x)$ is given in (\ref{gct}), and $A_\mu=A^a_\mu T_a$.
This transformation is to correspond with that the connection
taken in two different local neighbourhoods coincides together in
an intersection region. We can see that the transformation rule
(\ref{gautrnI}) is to combine of both the $4$-dimensional external
coordinate and the usual gauge transformations.\footnote{It is
also interesting to note that as analyzed in references
\cite{Jackiw1978,Jackiw1980,JackiwPi2002} the coordinate
transformation can be combined with the gauge transformation to
form a \emph{gauge-covariant coordinate transformation}.} This is
due to that it has an origin from the local coordinate
transformations of the higher dimensional space-time. It can be
verified that an infinitesimal variation of $A^a_\mu$ is taken as
\begin{equation}
\delta A^a_\mu(x)=-\partial_\mu\epsilon^\nu(x)
A^a_\nu(x)-\epsilon^\nu(x)\partial_\nu A^a_\mu(x)+
f^a_{bc}\alpha^b(x)A^c_\mu(x)-\partial_\mu\alpha^a(x),\label{infgautr}
\end{equation}
where infinitesimal parameters $\epsilon^\mu(x)$ and $\alpha^a(x)$
are defined as, $x'^\mu=x^\mu+\epsilon^\mu(x)$ and
$\textrm{exp}\{i\alpha^a(x)T_a\}\cong 1+i\alpha^a(x)T_a$,
respectively. The first two terms in Eq. (\ref{infgautr}) come
from the infinitesimal transformation of the $4$-dimensional
external coordinates while the remaining ones present the usually
infinitesimal gauge transformation.

We have seen that the gauge fields $A^a_\mu$ gives the covariant
way to determine the directions of the usual $4$-dimensional world
in $B^D$. Since an arbitrary particle moving along the
$4$-dimensional external directions of the bulk may be coupled
with $A^a_\mu$. This is easily seen by the appearance of the gauge
fields $A^a_\mu$ in the derivatives $\hat{\partial}_\mu$ implying
that the field propagating in the bulk having non-zero derivatives
in the internal variables would interact to $A^a_\mu$. It is
interesting to interpret that corresponding gauge charges are
generated dynamically through the physical kinetics in the
internal spaces. This means that such interaction is of the
manifestation of properties of the internal dynamics. In addition
to such coupling perspective, there will be known the coupling
occurring in the context of the usual gauge theory in which the
gauge fields $A^a_\mu$ are coupled to the fields of carrying out
non-trivial representations of the group $G$, as we will discuss
in Sec. \ref{gaugecharges}. Eventually, the connection one-form is
realized as the elegant object in the mathematical point of view
while the gauge fields $A^a_\mu$ carry out the interesting
physical significance as force-carrying particles.

Any vector field $V(X)$ on $B^D$ is always written in external and
internal components in the following form
\begin{eqnarray}
V(X)=V^\mu_H(X)\hat{\partial}_\mu+V^a_V(X)\hat{\partial}_a.
\end{eqnarray}
Since $V^\mu_H(X)$ and $V^a_V(X)$ do not mix with each other, they
are thought of as the independently physical fields at the
fundamentally higher dimensional level. It is also straightforward
to generalize that for a general tensor field on the bulk in which
there will have additionally hybrid fields including both the
external and internal indices.

A better understanding about the topological non-triviality of the
bundle $B^D$ is given through a curvature two-form $\Omega$ of the
given connection. The curvature would tell us how to distinguish a
given connections on $B^D$ that is sufficiently different to the
flat connection in which its curvature vanishes to correspond with
the trivial bundle $B^D$. This means that it measures the twisting
of the bundle $B^D$. The curvature is defined by the Cartan's
structure equation as
\begin{equation}
\Omega=d\omega+\omega\wedge\omega,\label{Cstreq}
\end{equation}
where $\wedge$ is the wedge product on the space of the
differential forms living the bulk. From this definition, one can
show that $\Omega$ acts on any pair of horizontal vectors to yield
an element of $\mathfrak{g}$ but vanishing for the remaining
cases. In other works, only the horizontal components of the
curvature are not zero in which they are defined explicitly as
follows
\begin{equation}
\Omega_{\mu\nu}(X)=ig_{\textrm{i}}e^{-i\theta}F_{\mu\nu}(x)e^{i\theta},
\end{equation}
where
\begin{equation}
F_{\mu\nu}(x)=\left(\partial_\mu A^a_\nu -\partial_\nu
A^a_\mu-g_{\textrm{i}}f^a_{bc}A^b_\mu A^c_\nu\right)T_a,
\end{equation}
is well-known as the Yang-Mills field strength tensor. In an
overlap region between any two neighbourhoods, the curvature
$\Omega$ has to obey, $\Omega=\Omega'$, leading to the rule of the
gauge transformation for $F_{\mu\nu}$ as follows
\begin{equation}
F'_{\mu\nu}(x')=\frac{\partial x^\rho}{\partial
x'^\mu}\frac{\partial x^\lambda}{\partial
x'^\nu}h(x)F_{\rho\lambda}(x)h^{-1}(x),
\end{equation}
where $h(x)$ is given in (\ref{gct}).

\subsection{\label{fundvectfield} Global frame of the vertical tangent spaces}
The Lie group $G$ acts smoothly and freely on $B^D$ as the active
diffeomorphism transformations,\footnote{The coordinates of two
points on $B^D$ are related together under the action of an
element $g\in G$, denoted $R_g$, given by \begin{equation}
X^M=(x^\mu,\hat{\theta}^a)\stackrel{R_g}{\longrightarrow}X'^M=(x^\mu,\hat{\theta}'^a),\label{rightac}
\end{equation}
where $\exp\{i\theta'^aT_a\}=\exp\{i\theta^aT_a\}g$. It is
important that this action occurs in the extremely natural way
because the definition (\ref{rightac}) is independent of choosing
of the local coordinate system as well as does not depend of
deforming of the bulk under the presence of matter sources.} known
as the right action, to induce naturally so-called fundamental
vector fields which are defined globally on it. All of them form a
vector space preserving the structure of the Lie algebra
$\mathfrak{g}$, thus, it is isomorphic to $\mathfrak{g}$. The
basic vector fields of this vector space denoted by $K_i(X)$
$(\equiv\hat{\partial}_i)$, with $i=1,2,...,D-4$, satisfy the
following Lie bracket
\begin{equation}
[K_i(X),K_j(X)]=\overline{f}^k_{ij}K_k(X),\label{KeqI}
\end{equation}
where $\overline{f}^k_{ij}=\Lambda f^k_{ij}$ are the corresponding
structure constants. These structure constants can also realized
as a tensor of type $(1,2)$ transforming under the global rotation
of the frame fields $K_i$ read
\begin{equation}
K_i(X)\longrightarrow K'_i(X)={A_i}^jK_j(X),\label{Krotation}
\end{equation}
where ${A_i}^j$ is in general a constant matrix of the general
linear group $GL(D-4,\mathbb{R})$. We should note that each
vertical tangent space is also isomorphic to $\mathfrak{g}$ due to
that the internal spaces are to deform smoothly of $G$. It is thus
very important that $\{K_i\}$ constitutes a global frame for the
whole vertical tangent spaces on $B^D$.

These frame fields are written explicitly in a local neighbourhood
as
\begin{equation}
K_i(X)={K_i}^a(X)\hat{\partial}_a.
\end{equation}
Here ${K_i}^a(X)$ is dimensionless and transforms as a covector
under the rule of the fibre coordinate transformation and as a
vector under the transformation (\ref{Krotation}). The fact that a
horizontal vector field is transported along a flow generated by a
fundamental vector field remains again another one. On the other
hand, we must impose a condition on the following Lie derivatives,
$[V_H,K_i(X)]\in H_XB^D$ for all of $i=1,...,D-4$, with $V_H$ to
be a horizontal vector field. Thus, this results in the equations
for ${K_i}^a(X)$ as
\begin{equation}
\left(\partial_\mu-g_{\textrm{i}}A^b_\mu\partial
_b\right){K_i}^a(X)=0.\label{KeqII}
\end{equation}
These first-order partial differential equations may be considered
as the equations describing the evolution of the frame fields
$K_i(X)$ under the dynamics of the bulk space-time. It is
important to be emphasized that these equations are properly
defined in a pure mathematical way rather than from the equations
of motion obtained by varying the corresponding action. This means
that the fields $K_i(X)$ would be not realized as the physical
degrees of freedom in the usual sense, but are the mathematically
axial fields.

The fundamental vector fields are used to describe the
infinitesimal displacement behaviour of the action of the Lie
group $G$ on $B^D$. It is expressed in terms of parameters of the
infinitesimal transformation $\varepsilon^i$ as
\begin{equation}
x'^\mu=x^\mu,\hspace*{1 cm}
\theta'^a=\theta^a+\varepsilon^i{K_i}^a.\label{Killdis}
\end{equation}
The flows of $K_i(X)$ running along the internal spaces lead to
the infinitesimal displacements obviously only performed along
these spaces. Since the $4$-dimensional external coordinates in
the same local neighbourhood do not change. Notice also that,
$\varepsilon^i$ are independent on the bulk coordinates because
the action of the Lie group $G$ is defined globally. The second
equation of Eq. (\ref{Killdis}) suggests that $K_i(X)$ play the
role of the infinitesimal generators corresponding with the active
diffeomorphism transformations resulting from the action of $G$ on
the bulk. With respect a physical field $\psi$ which is an active
diffeomorphism invariance under the action of $G$, then Lie
derivatives of $\psi$ along all fields $K_i$ vanish
\begin{equation}
\mathcal{L}_{K_i}\psi=0.\label{Gsymm}
\end{equation}
Here the Lie derivatives are defined for general mathematical
objects rather than only the usual tensor fields on $B^D$. The
vector fields are characterized by Eq. (\ref{Gsymm}) to point out
the symmetric directions with resepct to the field $\psi$ in the
internal spaces. In this case, the internal dynamics of $\psi$ is
governed by the \emph{intrinsic symmetry nature} under $G$ rather
than by the sources. As indicated in the paper
\cite{Gaul-Rovelli2000}, the active diffeomorphism invariance is a
property of the dynamical theory itself whereas the passive
diffeomorphism invariance is a property of the formulation of a
dynamical theory. A specially interesting case in which the
right-hand of Eq. (\ref{Gsymm}) is not vanished but is proportion
to $\psi$ will lead to the conformal diffeomorphism invariance for
$\psi$ under $G$.

Let us introduce dual fields of $K_i(X)$, denoted by $K^i(X)$,
satisfying the following conditions
\begin{equation}
\langle K^i,K_j\rangle=\delta^i_j,
\end{equation}
which form a global basis for the internal one-form fields on
$B^D$. Their local form is taken as
\begin{equation}
K^i(X)={K^i}_a(X)\left(d\hat{\theta}^a+\frac{g_{\textrm{i}}A^a_\mu
dx^\mu}{\Lambda}\right),
\end{equation}
where ${K^i}_a(X)$ transform as an vector under the rule of the
fibre coordinate transformation and as covector under the rotation
of $K_i$ given in (\ref{Krotation}). With the help of both
(\ref{KeqI}) and (\ref{KeqII}) equations, it can be easily
verified that $K^i(X)$ obey the equations extended from the usual
Maurer-Cartan's structure equations as follows
\begin{equation}
dK^i=-\frac{1}{2}\overline{f}^i_{jk}K^j\wedge
K^k+\frac{g_{\textrm{i}}}{2\Lambda}F^i_{\mu\nu}dx^\mu\wedge
dx^\nu,\label{ExMau-Careqs}
\end{equation}
where $F^i_{\mu\nu}$ is defined in terms of
$A^i_\mu={K^i}_aA^a_\mu$ of the following form
\begin{equation}
F^i_{\mu\nu}=\partial_\mu A^i_\nu-\partial_\nu
A^i_\mu+g_{\textrm{i}}f^i_{jk}A^j_\mu A^k_\nu.
\end{equation}
The mathematical interpretation of $F^i_{\mu\nu}$ is to measure
the failure of integrability of the horizontal distribution
because the Lie bracket of any two horizontal frame fields is
given as
\begin{equation}
[\hat{\partial}_\mu,\hat{\partial_\nu}]=-\frac{g_{\textrm{i}}}{\Lambda}F^i_{\mu\nu}K_i,
\end{equation}
which leads to a vertical vector. Because of that the bundle $B^D$
is topologically non-trivial, the external submanifolds are in
general joint together. In the case of all the components
$F^i_{\mu\nu}$ vanishing, the horizontal distribution is
integrable since by the Frobenius's theorem the higher dimensional
space-time $B^D$ is also foliated by the $4$-dimensional external
submanifolds.

\section{\label{bulkgra} Bulk Gravity}
In the minimal formalism, the fundamental variables describing the
dynamics of the gravity consist of the gauge fields $A^a_\mu$
together with the geometrical field (a bulk metric tensor)
determining the smooth of the bulk manifold as well as a causal
structure between the physical events.\footnote{In particular, new
degrees of freedom may occur in a remarkable way in the higher
dimensional extensions of the gravity. For example, these come
from the contorsion field which is considered even in four
dimensions. An alternative possible is of the ($p+1$)-form gauge
fields ($p>0$) \cite{Henneaux-Teitelboim1986} occurring almost in
supergravity theories whose sources are the charged objects in the
higher dimensional space-time, $p$-branes. One of these, the
rank-$2$ antisymmetric Kalb-Ramond (KR) field
\cite{Kalb-Ramond1974} was considered in the large
extra-dimensional \cite{Mukhopadhyaya_PRD2002} and Randall-Sundrum
\cite{Mukhopadhyaya2002&2004} compactification scenarios.} An
ansatz for the bulk metric is given locally in the most general
form as follows
\begin{equation}
ds^2_B=g_{\mu\nu}(X)dx^\mu dx^\nu-\gamma_{ij}(X)K^iK^j,\label{bmt}
\end{equation}
which describes two separately infinitesimal invariant intervals
on the external and internal spaces, respectively. The signature
of the metric $g_{\mu\nu}$ is chosen as, $(+---)$, since the sign
``$-$'' in the second term of Eq. (\ref{bmt}) refers the
space-like nature of the internal spaces. In this way,
$\gamma_{ij}$ includes obviously global degrees of freedom due to
the internal coframe fields $K^i$ defined globally on the bulk. It
can thus be interpreted in sense of the volume modulus fields
which set dynamically the size of the internal spaces. Under this
description, all of them are manifestly independent with each
other. Since they are realized as the fundamental fields unified
in the same non-trivial geometrical structure of the higher
dimensional space-time. In addition to the fields created the bulk
gravity, we also have the mathematically axial field ${K_i}^a$, or
its dual ${K^i}_a$.

It is interesting that the internal metric $\gamma_{ij}$ enjoys
naturally an important one which it is a constant matrix with
respect the coframe fields $K^i$. In this case, one may choose
others so that $\gamma_{ij}$ would be diagonal. It thus leads to
the metric $\gamma_{ij}$ that may be expressed as follows
\begin{equation}
\gamma_{ij}=v_i\delta_{ij},\label{cmim}
\end{equation}
where $v_i$ are all positive constants, and $\delta_{ij}$ is the
Kronecker delta function. Furthermore, the internal metric in
(\ref{cmim}) may also be re-scaled by an appropriately conformal
factor in order for one of the components $v_i$ to be equal of the
constant value, $1$. Thus, without loss of generality we can
choose $v_i=1$ for $i=D-4$. It is very important to note that the
above given conclusion would be incorrect if an manifold is
impossible to define the global coframe fields. This is because of
the metric in such case carrying out local degrees of freedom
which would be changed under the local coordinate transformation.
An interesting interpretation of $(D-5)$ undetermined values
$v_i$, $i=1,...D-5$, is that they should be fixed dynamically
corresponding with VEVs of the modulus fields. This can be thought
of as the physical compactification mechanism stabilizing the size
of the internal spaces. Correspondingly, in the present situation
we would like to study the dynamics of the internal metric with
the scheme to be as economical as possible in which the relevant
dynamic fields are given in the following form
\begin{equation}
\gamma_{ij}(X)=e^{\phi_i(X)}\delta_{ij},\label{ExGMI}
\end{equation}
due to the fact that $\gamma_{ij}$ is positive definition, here
$\phi_{D-4}=0$. The transformations given in Eq. (\ref{Krotation})
preserving the form (\ref{ExGMI}) correspond with taking a
dilation transformation on $\phi_i(X)$ with $i=1,...,D-5$, as
\begin{equation}
\phi_i(X)\longrightarrow\phi_i(X)+\lambda_i,
\end{equation}
where $\lambda_i$ are constants. For $i=1,...,D-5$, $v_i$ are
expected to relate with VEVs of the modulus fields
$\langle\phi_i\rangle$ as, $v_i=e^{\langle\phi_i\rangle}$.
Furthermore, we consider the modulus fields being independent of
the internal coordinates, $\phi_i=\phi_i(x)$, which correspond
with the metric on each internal space to be invariant under the
action of $G$. As we will see later on, the internal metric under
consideration is consistent with that at the ground state of the
bulk.

To evaluate the change of any tensor field on the bulk space-time
along a vector field, we need to introduce a linear connection to
construct an operator of the covariant derivative. It is
interesting that there exist many possible connections on the
foliated manifolds as given in \cite{Benjacu2006}. However, the
fact that connection will be considered in the present work is
torsion-free (or symmetric) and compatible with metric, also
well-known as the Levi-Civita connection. It is determined
uniquely on a (pseudo-) Riemannian manifold.\footnote{Dealing with
non-symmetric metric connection has received some interesting
results in four-dimensions as well as higher-dimensions in which
new degrees of freedom coming from contorsion field would be
included into.} In general, the coefficients of the considered
linear connection and components of the Riemann curvature tensor
are taken exactly in the terms of the bulk metric $G_{MN}$ and
non-holonomic functions $C^P_{MN}$ as
\begin{eqnarray}
\Gamma^P_{MN}&=&\frac{G^{PQ}}{2}\left(\partial_MG_{NQ}+\partial_NG_{MQ}-\partial_QG_{MN}\right)
+\frac{G^{PQ}}{2}\left(C^O_{QM}G_{ON}+C^O_{QN}G_{OM}\right)+\frac{C^P_{MN}}{2},
\nonumber \\
R^O_{MPN}&=&\partial_N[\Gamma^O_{PM}]-\partial_P[\Gamma^O_{NM}]+
\Gamma^Q_{PM}\Gamma^O_{NQ}-\Gamma^Q_{NM}\Gamma^O_{PQ}+C^Q_{PN}\Gamma^O_{QM}.
\end{eqnarray}
Note that, in the non-holonomic basis the torsion-free condition
of the metric connection does not result in the usual symmetric
condition, $\Gamma^P_{MN}=\Gamma^P_{NM}$. However, in the natural
or holonomic basis, ($\{\partial_\mu\},\{\hat{\partial}_a\}$), in
which all of the factors $C^P_{MN}$ vanish identically, then we
will get again the familiar expressions for $\Gamma^P_{MN}$ as
well as $R^O_{MPN}$. It is convenient to adapt the physical basic,
$(\{\hat{\partial}_\mu\},\{\hat{\partial}_i\})$, where explicit
expressions for $C^P_{MN}$ and $\Gamma^P_{MN}$ are given in the
appendix. It is important that with the Levi-Civita connection on
$B^D$ the covariant derivative of a horizontal vector in another
one would in general not result in a vector of the horizontal
tangent space, and similarly neither do vertical vectors.

For our setup, the classical dynamics of the pure bulk gravity is
governed by the action to be invariant under both the local
coordinate transformations (\ref{gct}) and the global
transformation (\ref{Krotation}) taking the form
\begin{equation}
S_{G}=M^{D-2}_{*}\int_{B^D}
d^DX\sqrt{|G|}\left[\frac{1}{2g^2_{\textrm{i}}M^2_*}\mathrm{Tr}\left(\Omega_{\mu\nu}\Omega^{\mu\nu}\right)+R-V(\phi_i)\right].\label{bgac}
\end{equation}
The fundamental Planck scale $M_{*}$ determines the scale of the
higher dimensional quantum gravity. $G=(-1)^{D-4}gK^2\gamma$, with
$K=\det({K^i}_a)$, is the determinant of the bulk metric expressed
through that of the external and internal metrics, denoted by $g$
and $\gamma$, respectively. The trace operator, $\mathrm{Tr}$,
refers to the non-degenerate inner product on the Lie algebra
$\mathfrak{g}$. The first term defines the dynamics of the gauge
fields $A^a_\mu $ which is naturally constructed in term of the
curvature of the connection at that
$\Omega^{\mu\nu}=g^{\mu\rho}g^{\nu\lambda}\Omega_{\rho\lambda}$
are the contravariant components of the curvature $\Omega$. The
scalar curvature $R$ of the bulk space-time at which the
space-time topology has been defined by the curvature $\Omega$ of
the principal bundle constructed as follows
\begin{equation}
R=G^{MN}R_{MN}=g^{\mu\nu}R^M_{\mu
M\nu}-\gamma^{ij}R^M_{iMj},\label{LRS}
\end{equation}
presents terms governing the dynamics of the external metric field
and the modulus fields. An explicit expansion of this term is
taken in the appendix. In the above given action, we have
introduced the potential for the modulus fields $V(\phi_i)$ that
can also be interpreted as an extension of $\phi_i$-dependence of
bulk cosmological constant.

The structure of the potential $V(\phi_i)$ constructed in such a
way that the indices of $\overline{f}^i_{jk}$, the internal metric
and its dual are contracted with each other has a remarkable form
as follows
\begin{eqnarray}
V(\phi_i)&=&\Lambda_B+\left(a_1\overline{f}^k_{il}\overline{f}^l_{kj}\gamma^{ij}
+a_2\overline{f}^p_{ik}\overline{f}^q_{jl}\gamma^{ij}\gamma^{kl}\gamma_{pq}\right)\nonumber \\
&=&\Lambda_B+\Lambda^2\left(a_1\sum_{i}f^k_{il}f^l_{ki}e^{-\phi_i}
+a_2\sum_{i,k,j}(f^k_{ij})^2e^{\phi_k-\phi_i-\phi_j}\right),\label{modPotI}
\end{eqnarray}
where $\Lambda_B$ is a bulk cosmological constant, and $a_{1,2}$
are dimensionless coupling constants. It is very important to see
that an essential point to give rise $V(\phi_i)$ is the structure
constants of Lie algebra $\mathfrak{g}$. In the usual way to
generate a potential for the metric field we may only contract the
indices of the metric to those of its dual without derivatives
leading to cosmological constant contribution. This means that the
above given potential would be a trivial constant with respect
internal manifolds that do not correspond to smooth deforming of
the non-Abelian Lie groups. Such manifolds have been most commonly
studied in the previously higher dimensional theories such as
$n$-torus $T^n$ or two-sphere $S^2$. Therefore, the presence of
this potential is thought of as one of the crucial natures of the
present framework. As will be shown below, $V(\phi_i)$ is also
added on other contributions coming from the bulk determinant and
the scalar curvature $R$ to generate a complete potential of the
moduli stabilization.

Using the explicit expression for $\Omega_{\mu\nu}$ derived in the
previous section and choosing a basis of $\mathfrak{g}$ so that,
$\mathrm{Tr}(T_aT_b)=\delta_{ab}/2$, the first term in the action
(\ref{bgac}) is then taken as
\begin{equation}
S^{1}_{G}=-\left(\frac{M_*}{\Lambda}\right)^{D-4}\int_{B^D}d^4xd^{D-4}\theta\sqrt{|G|}\frac{F^a_{\mu\nu}F^{a\mu\nu}}{4}.\label{YMaction}
\end{equation}
The expression under the integral is just the conventional
Yang-Mills Lagrangian but for the gauge fields $A^a_\mu$ lying on
the total space $B^D$ rather than on the basic space $M^4$. To get
the standard form, the factor $(M_*/\Lambda)^{D-4}$ should be
absorbed by normalizing the gauge fields $A^a_\mu$ which will lead
to new gauge coupling
\begin{equation}
g_{\textrm{i}}\left(\frac{\Lambda}{M_*}\right)^{\frac{D-4}{2}},\label{LahAI}
\end{equation}
that remains dimensionless.

Now let us be straightforward to the last two terms in the action
(\ref{bgac}). By explicitly expanding we get up to total
derivatives as
\begin{eqnarray}
S^{2}_{G}&=&\int
d^4xd^{D-4}\theta\sqrt{|g|}K\left[\overline{M}^2_Pe^{\widetilde{\phi}/\overline{M}_P}\mathcal{L}_g+\mathcal{L}_{\widetilde{\phi}}
-\frac{g^2_{\textrm{i}}}{4}\left(\frac{\overline{M}_P}{\Lambda}\right)^{2}\sum_ie^{(\widetilde{\phi}+\widetilde{\phi}_i)/\overline{M}_P}F^i_{\mu\nu}F^{i\mu\nu}\right],\label{EHa} \\
\mathcal{L}_g&=&\hat{R}+\frac{1}{4}\left(\hat{\partial}_ig^{\mu\nu}\hat{\partial}^ig_{\mu\nu}-g^{\mu\nu}g^{\rho\lambda}
\hat{\partial}_ig_{\mu\nu}\hat{\partial}^ig_{\rho\lambda}\right),\label{4Dgra}
\label{4Dgra}\\
\mathcal{L}_{\widetilde{\phi}}&=&
\frac{e^{\widetilde{\phi}/\overline{M}_P}}{2}\sum^{D-5}_{i}\hat{\partial}_\mu\widetilde{\phi}_{i}\hat{\partial}^\mu\widetilde{\phi}_{i}
-\widetilde{V}(\widetilde{\phi_i}).\label{modLag}
\end{eqnarray}
The contravariant partial derivatives are defined by
\begin{equation}
\hat{\partial}^\mu=g^{\mu\nu}\hat{\partial}_\nu, \  \
\hat{\partial}^i=\gamma^{ij}\hat{\partial}_j=e^{-\widetilde{\phi_i}/\overline{M}_P}\hat{\partial}_i,
\end{equation}
with $\overline{M}^2_P=M^{D-2}_*/\Lambda^{D-4}$. We have
normalized the modulus fields as
\begin{equation}
\widetilde{\phi_i}=\overline{M}_P\phi_i,
\end{equation}
which have the dimension of mass, and
$\widetilde{\phi}=\sum_i\widetilde{\phi_i}/2$. The conformal
factor $e^{\widetilde{\phi}/\overline{M}_P}$ defines volume of the
internal spaces. The dynamics of $g_{\mu\nu}$ and
$\widetilde{\phi}_i$ in $B^D$ is determined by $\mathcal{L}_g$ and
$\mathcal{L}_{\widetilde{\phi}}$, respectively. The last term in
the action (\ref{EHa}) plays the coupling role. The
$4$-dimensional standard scalar curvature $\hat{R}$ is given to
associate with the $4$-dimensional external metric $g_{\mu\nu}$ as
\begin{equation}
\hat{R}=g^{\mu\nu}\left(\hat{\partial}_\nu\Gamma^\lambda_{\lambda\mu}-\hat{\partial}_\lambda\Gamma^\lambda_{\nu\mu}+
\Gamma^\rho_{\lambda\mu}\Gamma^\lambda_{\nu\rho}-\Gamma^\rho_{\nu\mu}\Gamma^\lambda_{\lambda\rho}\right),\label{4Dscur}
\end{equation}
where $\Gamma^\lambda_{\mu\nu}$ are the horizontal components of
the Christoffel symbols $\Gamma^P_{MN}$. Therefore, it consists of
two parts at which one presents the external kinetic energy term
for $g_{\mu\nu}$ or the conventional $4$-dimensional
Einstein-Hilbert term, and another presents the coupling term
between $g_{\mu\nu}$ and the gauge fields $A^a_\mu$. The potential
$\widetilde{V}(\widetilde{\phi}_i)$ is taken as
\begin{equation}
\widetilde{V}=(\Lambda\overline{M}_P)^2e^{\widetilde{\phi}/\overline{M}_P}\left[\frac{\Lambda_B}{\Lambda^2}+\left(a_1-\frac{1}{2}\right)\sum_{i}f^k_{il}f^l_{ki}
e^{-\widetilde{\phi}_i/\overline{M}_P}+\left(a_2+\frac{1}{4}\right)\sum_{i,k,j}(f^k_{ij})^2e^{(\widetilde{\phi}_k-\widetilde{\phi}_i-\widetilde{\phi}_j)/\overline{M}_P}\right].\label{cpot}
\end{equation}
Note that the terms corresponding to the factors, $-1/2$ and
$1/4$, in parentheses are to come from the scalar curvature $R$.

It is quite clearly that the potential
$\widetilde{V}(\widetilde{\phi}_i)$ will be a promise of moduli
stabilization at fundamental level. This achieves only if
$\widetilde{V}(\widetilde{\phi}_i)$ holds a stable minimum
associated with that VEVs of the modulus fields are no longer
chosen in an arbitrary way but are completely determined. On the
other hand, the size of the internal spaces is dynamically fixed.
In contrast, there will occur corresponding massless scalar fields
in the effective low energy theory which cause dangerous
phenomenological problems.\footnote{In such case, an effective
potential for the modulus fields has to be generated in various
ways. It has been shown that such potential can arise via quantum
effects of pure geometrical and non-geometrical fields
\cite{Appelquist-Chodos1983,Rubin-RothPLB1983,Ponton-Poppitz2001}.
The dynamics of a bulk scalar field as proposed in
\cite{Goldberger-WisePRL1999} also provides a mechanism for
stabilizing the size of the extra dimension in the Randall-Sundrum
model. An alternative possibility is to come from the gaugino
condensation \cite{Derendinger1985,Dine1985} which is also
considered to break dynamically the supersymmetry
\cite{Nilles2004}} Even though the modulus fields are stabilized
by $\widetilde{V}(\widetilde{\phi}_i)$, Kaluza-Klein spectrum of
an arbitrary particle propagating in the vacuum background of the
modulus fields would still be unspecified by corresponding Laplace
operator unknown. In fact, we have to finish a last step by
solving Eq. (\ref{KeqII}) or Eq. (\ref{ExMau-Careqs}) to determine
the internal partial derivatives $\hat{\partial}_i$. However, in
the low energy regime in which the twisting of the bundle $B^D$ is
essentially trivial, since this is done in a much more simple way.
It is due to that the internal frame fields $\hat{\partial}_i$ and
their dual $K^i(x,\theta)$ may be approximately taken as those of
the left-invariant vector fields and one-forms, respectively, on
the Lie group $G$.

\begin{figure}[t]
 \centering
\begin{tabular}{cc}
\includegraphics[width=0.45 \textwidth]{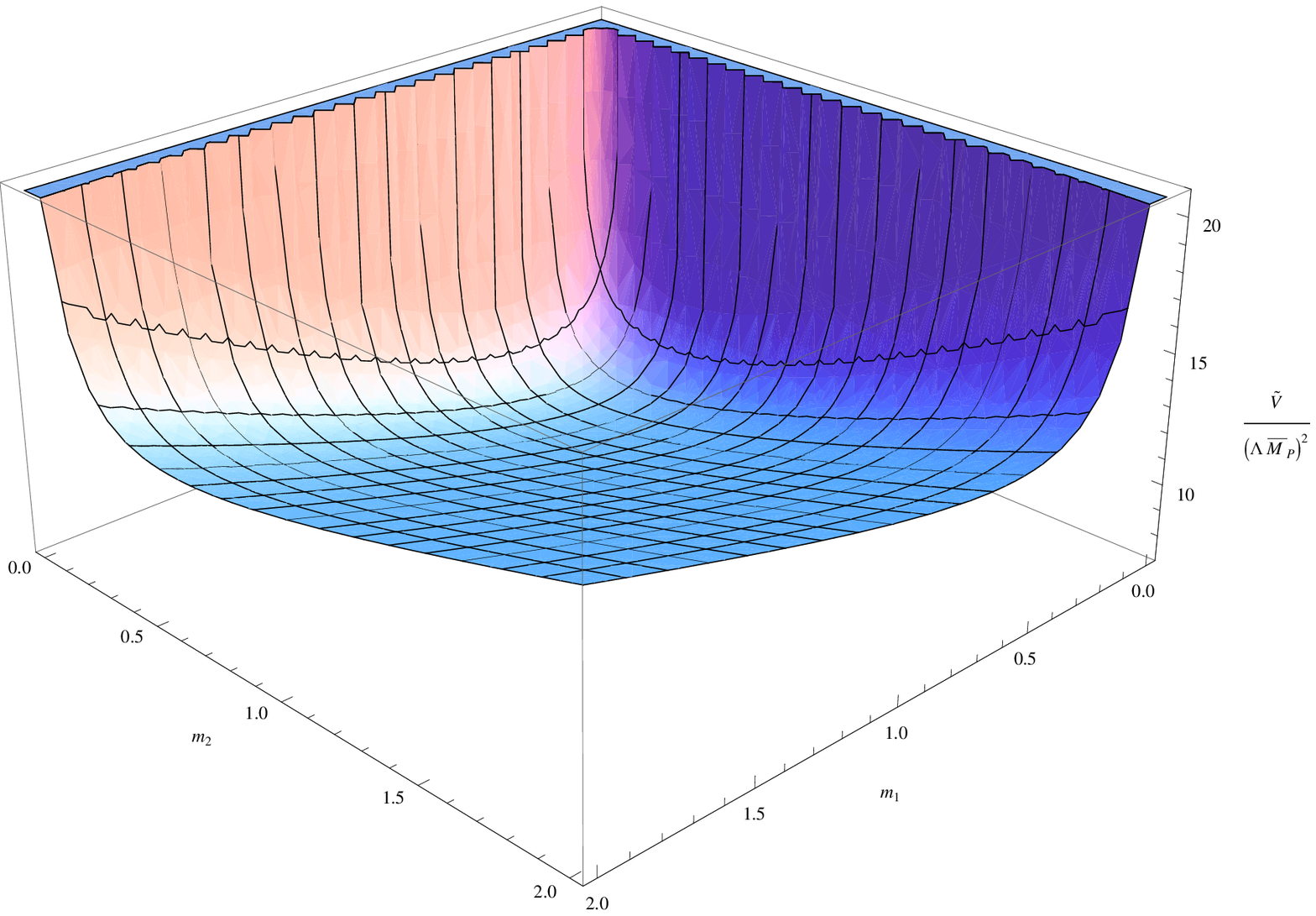}
\hspace*{0.05\textwidth}
\includegraphics[width=0.35 \textwidth]{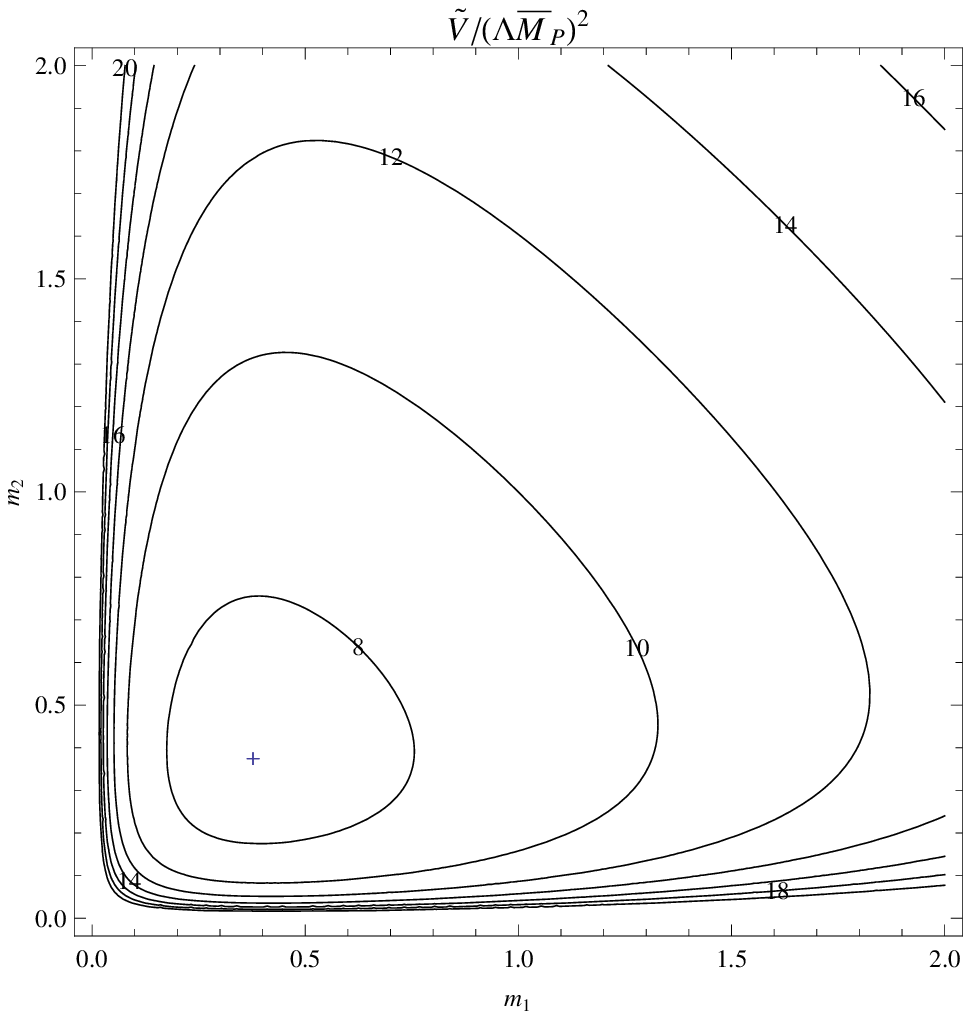}
\end{tabular}
  \caption{The shape of the potential
  $\widetilde{V}(\widetilde{\phi}_i)$, defined in unit of
  $(\Lambda\overline{M}_P)^2$, is plotted as the function, for convenience, of $m_i\equiv e^{\widetilde{\phi}_i/\overline{M}_P}$ ($i=1,2$) in the
  case of the internal manifolds being diffeomorphic to the Lie group $\mathrm{SU}(2)$. The parameters of the potential are valued as
  $\frac{\Lambda_B}{\Lambda^2}=5$, $a_1=0$, and $a_2=\frac{3}{4}$.
Left panel: The three-dimensional plot of the given potential has
an absolute minimum seen quite clearly. Right panel: The contour
plot of the given potential with ``+'' referring the absolute
minimum.}\label{modpoSU2}
\end{figure}

In order to verify the existence of the stable minimum with
respect the moduli stabilization potential
$\widetilde{V}(\widetilde{\phi}_i)$ above, let us consider a
simple illustration example of the internal spaces which are
smoothly equivalent to the group manifold $\mathrm{SU}(2)$. It is
well known that this manifold is diffeomorphic to the $3$-sphere.
For this case, it is not difficult to find the corresponding
potential for two modulus fields in which some conditions on the
parameters of the potential are required to guarantee it being
bounded from below. Its form is shown, for example, in FIG.
\ref{modpoSU2} with de Sitter vacuum. Note that, in the realistic
theory the minimum value of the moduli stabilization potential has
to be fine tuned in the way of total vacuum energy density being
consistent with the cosmological constant
\cite{Weinberg1989,Bousso2008}. We would like to emphasize that
this illustration may also be true for more complex cases of the
special unitary group as well as the special orthogonal group.

It is important that $\widetilde{V}(\widetilde{\phi}_i)$ is
emerged from the intrinsic properties of the higher dimensional
space-time. Therefore, the present work may provide the natural
mechanism to understand the dynamics of the extra dimensional
compactification. Notice that the classical potential of the
volume modulus fields can also get effective contributions from
others. It is interesting to think of that the inflation can be
explained in the framework in connection with the moduli
stabilization attracted in much of the previous works, for
example, see in \cite{radion-inflation}.

Let us end this section with a few important comments on the
construction of the higher dimensional gravitational interaction
above. According to this description we can see that the higher
dimensional gravity includes the gauge interaction of the
symmetric group $G$ mediated by the gauge fields $A^a_\mu$. They
are fundamental dynamic variables which are mathematically
characterized to the space-time at which they define the
topological non-triviality of $B^D$ via the Yang-Mills field
strength tensor $F_{\mu\nu}$. Thus, their dynamics is to
correspond with the topological dynamics of the space-time. This
is clearly to be different with the usual gauge theories at which
the gauge fields lay on the space-time of which is a basic
manifold of a principal bundle, and the corresponding dynamics
causes to the curvature of the space-time described by the
dynamical metric field. Since one should not confuse this
framework with that the ordinary gauge theory is coupled to the
Einstein gravity and the scalars. This proposal is also completely
different with the Kaluza-Klein theories. It is due to that in
these theories at low energy the usual gauge fields are identified
as part of the higher-dimensional metric via the Kaluza-Klein
reduction. It is interesting to see that the gravity mediated by
$A^a_\mu$ leads to the consistent quantum description. This means
that the quantum behavior of the space-time can be understood
manifestly by the quantum dynamics of the gauge fields $A^a_\mu$.
A detailed study will be published elsewhere. Furthermore, this
framework can also help better understanding about the
gauge/gravity correspondence
\cite{Maldacena98&99,Gubsere-Klebanov-Polyakov1998,Witten1998}
which relates the weak coupled classical gravity to the strong
coupled field theory in a flat background in one lower dimension.
So far, this is only conjectured to base on studying of the string
theory in which the black $p$-brane picture
\cite{Horowitz-Strominger1991} in certain limit is dual to the
D$p$-brane picture obtained in the weak string coupling limit of
the correponding superstring theory.

\section{Ground state of Bulk and perturbative gravity}
\subsection{\label{BGvac}Vacuum solution of $B^D$}
Let us look for what means actually by classical vacuum state of
$B^D$. In the classical ground state, the matter sources are
absent since the twisting of the higher dimensional bulk is
obviously trivial. This means that there is a canonical splitting
of $B^D$ into the $4$-dimensional space-time and the
($D-4$)-dimensional extra space without the gauge fields
$A^a_\mu$. On other the hand, we may consider a solution
describing the ground state geometry of the bulk with the topology
given as
\begin{equation}
B^D=M^4\times G,\label{vama}
\end{equation}
with the corresponding metric read
\begin{equation}
G^0_{MN}=\left(%
\begin{array}{cc}
  g_{\mu\nu}(x) & 0 \\
  0 & e^{\langle\widetilde{\phi}_i\rangle/\overline{M}_P}\delta_{ij} \\
\end{array}%
\right).
\end{equation}
Notice that, the ground state of $B^D$ must admit the internal
metric associated with the stable minimum of the potential
$\widetilde{V}(\widetilde{\phi}_i)$ in which the modulus fields
get VEVs denoted by $\langle\widetilde{\phi}_i\rangle$ (with
$i=1,...,D-5$), and $\langle\widetilde{\phi}_{D-4}\rangle=0$.
Indeed, this is applicable for the geometry of the ground state
because $K_i$ (or $\hat{\partial}_i$) and $K^i$ now correspond to
the basic fields of the left-invariant vector fields and one-forms
on the Lie group $G$. In this case, it is interesting to result in
the left-invariant metric on $G$. The above topological product
implies that the vacuum geometry of the bulk is foliated by either
$G$ or $M^4$.

By the perfect feature of the ground state geometry, we are ready
to write the Einstein field equation for the external metric
$g_{\mu\nu}(x)$ as follows
\begin{equation}
\hat{R}_{\mu\nu}-\frac{1}{2}g_{\mu\nu}(\hat{R}-\Lambda_{eff})=0\label{vEeq}.
\end{equation}
Here $\hat{R}_{\mu\nu}$ and $\hat{R}$ are the Ricci and scalar
curvatures, respectively, of the manifold $M^4$. The effective
cosmological constant $\Lambda_{eff}$ is given by
\begin{equation}
\Lambda_{eff}=\frac{e^{-\langle\widetilde{\phi}\rangle/\overline{M}_P}\widetilde{V}_{\textrm{min}}}{\overline{M}^2_P},
\end{equation}
where $\widetilde{V}_{\textrm{min}}$ is the value of the potential
$\widetilde{V}(\widetilde{\phi}_i)$ at its absolute minimum, and
$\langle\widetilde{\phi}\rangle=\sum_i\langle\widetilde{\phi_i}\rangle/2$.
The solution for Eq. (\ref{vEeq}) can be derived by the Ricci
tensor satisfying in the following relation
\begin{equation}
\hat{R}_{\mu\nu}=\frac{\Lambda_{eff}}{2}g_{\mu\nu}.\label{vasol}
\end{equation}
This solution results in that the manifold $M^4$ is the
$4$-dimensional Einstein space which has maximal symmetry. If
$\Lambda_{eff}$ is non-zero, depending on its sign to be negative
or positive, then $M^4$ is of the Anti de Sitter or the de Sitter
space, respectively. With respect the remaining case that
$\widetilde{V}_{\textrm{min}}$ vanishes where the parameters of
the potential must be finely-turned to the contributions in the
potential exactly canceled with each other, we have the solution
of the Ricci-flat $4$-dimensional manifold. One of such ones is
the $4$-dimensional Minkowski-flat space-time which has also to
fulfil the stronger condition that all the components of the
Riemannian tensor vanish identically. At classical level, the
Minkowski space-time which is stable by the positive energy
theorem \cite{WittenCMP1981} has the lowest energy. Therefore, the
classical ground state for $B^D$ is determined by the geometry
$\mathbb{R}^{1,3}\times G$ whose isometry group is the product of
the $4$-dimensional Poincar\'{e} group and the Lie group $G$.

So far we have concentrated mostly to discuss the ground state of
$B^D$ in the classical level. However, a question can be asked
whether the vacuum state found above is stable or modified under
the quantum effects. In the following, we would like to comment
briefly about a true vacuum solution for $B^D$ in point of view of
the quantum field theory in which the non-perturbative effects
would reveal the structure of this vacuum. However, a more
detailed study for such is beyond the scope of the present work
but will be done in our feature work. It is important to see that
the true vacuum geometry of $B^D$ has an extremely complicated
structure due to the non-trivial topology in the space of the
gauge fields. It is defined as a linear combination of
topologically inequivalent energy-degenerate vacua. Each such
vacuum is corresponded to a \emph{instanton number} $n$ which is a
geometric invariant characterizing the non-trivial topology of the
bundle $B^D$. This quantum number is derived by integrating the
second Chern character defined on the total space over a
four-dimensional external manifold.\footnote{These numbers are not
an integer in general unless the internal spaces are smooth copies
of the Lie group $\mathrm{SU}(N)$.} The classical ground state has
been looked for above to only correspond with the case of the
trivial instanton number, $n=0$. On the other hand, it is
inconsistent with true vacuum of $B^D$. Classically, the system
obviously lies at the rest at a ground state with defined quantum
number. Quantum-mechanically, zero-point fluctuations would play
the important role which leads to the tunneling among different
topological vacuum solutions, also well known as the instanton
effects. Therefore, the geometry of the ground state for $B^D$ is
precisely a superposition of these solutions presenting the
quantum nature of the space-time. In this case there will not be
simple to take the gauge fields $A^a_\mu$ to be zero and the
globally flat metric with respect the external spaces as performed
above. The instanton effects are thus expected to make some
interestingly physical implications, such as in cosmology, as well
as modify the classical properties of theory. Notice that, we have
not been interesting in the quantum effects on the classical
potential $\widetilde{V}(\widetilde{\phi}_i)$ in the above given
analysis.

\subsection{\label{BGD}Effective dynamics of the bulk gravitation below the
compactification scale}

The aim of this subsection is to study the physical particle
spectrum of the bulk gravity in the effective low energy theory
beneath the compactification scale in which we will mainly focus
the external metric field and the modulus fields. As usual, the
low energy degrees of freedom of the higher dimensional gravity
are obtained by the decomposition of the higher dimensional
massless graviton via the KK dimensional reduction. The result
includes, in the $4$-dimensional viewpoint, the massless zero
modes of graviton, vectors, radion (trace of the internal metric),
scalars (traceless of the internal metric), and their KK
excitations \cite{Giudice1999,Han1999}. In the case of the
internal space to be an orbifold, the zero mode of vectors will be
eliminated due to the orbifold projection. For the present work,
the low energy investigation of the bulk gravity may be done in a
simple way. The effective degrees of freedom consist of the
massive gauge bosons $A^a_\mu$ whose masses are generated by the
Higgs mechanism that will be discussed in the later section and
slightly perturbative fields around the classical background
solution given above.

In the fact, the present observed universe is essentially flat
meaning that the external spaces may be approximately considered
as the $4$-dimensional Minkowski space-time. Furthermore, because
of our present context which is a braneless scenario, this
background is not broken by the presence of branes as their
surface tension exceeds the fourth power of the fundamental Plank
scale, $M^4_{*}$ \cite{Sundrum1999b}. Since in the low energy
limit we can investigate the physical progress on the bulk
geometry to correspond with the above classical background.

In the weak field limit, the effective dynamics of the external
metric field and the modulus fields may be studied by expanding
them perturbatively around the ground state metric at the leading
order as
\begin{equation}
g_{\mu\nu}=\eta_{\mu\nu}+M^{-1}_\textrm{Pl}h_{\mu\nu},\ \
\widetilde{\phi}_i=\langle\widetilde{\phi}_i\rangle+\langle
V_G\rangle^{-\frac{1}{2}}\widetilde{h}_i, \  \ i\neq
D-4,\label{Gravityexp}
\end{equation}
where $\eta_{\mu\nu}$ is the $4$-dimensional Minkowski background.
The effective $4$-dimensional Plank scale $M_\textrm{Pl}$ and the
vacuum volume of the internal spaces $\langle V_G\rangle$ are
determined immediately via the fundamental parameters as follows
\begin{equation}
M^2_\textrm{Pl}=\frac{M^{D-2}_*\langle
V_G\rangle}{\Lambda^{D-4}},\ \ \langle
V_G\rangle=\int_{G}d^{D-4}\theta
K(\theta)e^{\langle\widetilde{\phi}\rangle/\overline{M}_P}.
\label{4DPl}
\end{equation}
It is very important to note that the first equality of
(\ref{4DPl}) is the usual relation to solve the hierarchy problem
in the large extra-dimensional model \cite{ADDmodel} with a
$n$-torus of equal radii $R$ in which $\langle
V_G\rangle/\Lambda^{D-4}\sim R^{D-4}$ and $M_*\sim 1\textrm{TeV}$.
Notice that, the specific energy scale $\Lambda$ of the compact
internal spaces in this framework should be so much larger than
the inverse radii of the extra dimensions in the large
extra-dimensional model. This is because all of particles are free
to propagate in the whole bulk. Since such extra dimensions would
not just play much the role of understanding the hierarchy problem
between the weak and the Plank scales in the observed world. The
$h_{\mu\nu}$ and $\widetilde{h}_i$ ($i\neq D-4$) express the small
curvature fluctuations of the external metric and modulus fields,
respectively. This means that both of them have to satisfy the
condition $|h_{\mu\nu}|, |\widetilde{h}_i|\ll 1$. In the above
given definition, these fields have been normalized so that they
have the canonical kinetic term and the mass dimension. Notice
that, $h_{\mu\nu}$ is the symmetric rank-$2$ tensor of ten
independent components which is of the reducible representation of
the $4$-dimensional Lorentz group. Thus it can be split into
irreducible ones in such a way as, $2\oplus 1\oplus 0\oplus 0$.
However, the fact that some degrees of freedom are consistently
eliminated by constraint equations leads to the physical degrees
of freedom to be less than original ones.

An effective Lagrangian determining the free evolution of
$h_{\mu\nu}$ in the bulk can be obtained by inserting the first
expansion of (\ref{Gravityexp}) into the Lagrangian (\ref{4Dgra}).
Then we find
\begin{equation}
\mathcal{L}_{h}=\frac{1}{2}\partial^\lambda
h^{\mu\nu}\partial_\lambda h_{\mu\nu}-\partial^\lambda
h^{\mu\nu}\partial_\mu h_{\lambda\nu}+\partial^\mu h\partial^\nu
h_{\mu\nu}-\frac{1}{2}\partial^\mu h\partial_\mu
h-\frac{1}{2}\left(
h^{\mu\nu}\square_{G}h_{\mu\nu}+h\square_{G}h\right),\label{E4Dac}
\end{equation}
where $h^{\mu\nu}=\eta^{\mu\lambda}\eta^{\nu\rho}h_{\lambda\rho}$,
$h=\eta^{\mu\nu}h_{\mu\nu}$,
$\partial^\mu=\eta^{\mu\nu}\partial_\nu$, and $\square_{G}$ is the
Laplace operator on the fixed internal spaces given as
\begin{equation}
\square_{G}=-\sum_{i}e^{-\langle\widetilde{\phi}_i\rangle/\overline{M}_P}{K_i}^a(\theta){K_i}^b(\theta)\hat{\partial}_a\hat{\partial}_b.\label{4DGee}
\end{equation}
By the compact topology of the internal spaces, the eigenvalues of
$\square_{G}$ are nonnegative and discrete. The equations of
motion derived by varying $\mathcal{L}_{h}$ with respect to the
perturbation $h_{\mu\nu}$ is of the form
\begin{equation}
\square h_{\mu\nu}-\partial_\lambda\partial_\mu
h^\lambda_\nu-\partial_\lambda\partial_\nu
h^\lambda_\mu+\eta_{\mu\nu}\partial_\lambda\partial_\sigma
h^{\lambda\sigma}+\partial_\mu\partial_\nu h-\eta_{\mu\nu}\square
h+\square_{G}\left(h_{\mu\nu}+\eta_{\mu\nu}h\right)=0.\label{E4Deq}
\end{equation}
We can show that both the Lagrangian (\ref{E4Dac}) and Eq.
(\ref{E4Deq}) are invariant under the local gauge transformation
\begin{equation}
h_{\mu\nu}(x,\hat{\theta})\longrightarrow
h_{\mu\nu}(x,\hat{\theta})-\partial_\mu\varepsilon_\nu(x)-\partial_\nu\varepsilon_\mu(x),\label{TengauTr}
\end{equation}
which is induced from the local external coordinate
transformation. However, it is very important to see that this
gauge transformation is not able to fix a gauge to eliminate some
unphysical degrees of freedom. This is due to the fact that the
field $h_{\mu\nu}$ is in general to depend on both the external
and the internal coordinates while the gauge transformation
parameters only depend on the external coordinates. It is
interestingly equivalent to the interpretation that under the
above gauge transformation $h_{\mu\nu}$ transforms locally on the
external spaces but globally on the internal ones. The global
transformation is then impossible to fix gauge hence in usual
sense the transformation (\ref{TengauTr}) does not generate
literally the local gauge transformation. On the contrary, it
would remove eight degrees of freedom to result in two physical
ones for $h_{\mu\nu}$. This is of the case that the field
$h_{\mu\nu}$ is independent on the internal coordinates.

The small fluctuations of the $4$-dimensional external metric can
be KK expanded as
\begin{equation}
h_{\mu\nu}(X)=\sum_nh^{(n)}_{\mu\nu}(x)Y^{(n)}(\hat{\theta}),
\end{equation}
where $Y^{(n)}(\hat{\theta})$ are the eigenvector of the operator
$\square_G$ corresponding to the eigenvalue $\lambda^2_n$, with
$n$ denoted to a set of quantum numbers characterizing to each
eigenvector. Each KK mode $h^{(n)}_{\mu\nu}(x)$ satisfies the
equations of motion in analogy with the Fierz-Pauli ones for the
massive graviton but the mass term in this equation replaced
by\footnote{Notice that the coefficient, $-1$, between
$h_{\mu\nu}$ and $\eta_{\mu\nu}h$ in the mass term of the
Fierz-Pauli equations is tuned by hand to have the consistence
theory. It not enforced by any symmetric property of theory
\cite{Hinterbichler2012}. For the present the coefficient, $+1$,
in the mass term given in (\ref{KKGramass}) is enforced by the
space-like feature of the internal kinetic term.}
\begin{equation}
\lambda^2_n\left(h^{(n)}_{\mu\nu}+\eta_{\mu\nu}h^{(n)}\right).\label{KKGramass}
\end{equation}
An important observation is that such mass terms clearly violate
the Fierz-Pauli tuning excepting $\lambda^2_0=0$ corresponding to
the massless zero mode which carries out two physical degrees of
freedom. With respect to every KK excitation, we only find the
constraint equations, $\partial^\mu h^{(n)}_{\mu\nu}+\partial_\nu
h^{(n)}=0$, which would remove the four degrees of freedom. The
constraint on the trace, $h^{(n)}=0$, is not able to be derived
unless the sign of the last term in (\ref{KKGramass}) is changed
oppositely. Thus, each KK excited mode propagating in the
effective $4$-dimensional Minkowski space-time carries exactly out
six physical degrees of freedom. It is very important that this
gives rise a $4$-dimensional massive graviton of five physical
degrees of freedom and a real scalar ghost of instability (a
scalar field with negative kinetic energy) which breaks the
unitarity of theory.

The linearization of the external metric field around the
classical background of $B^D$ yields to the unstable massive
scalar excitations in the effective $4$-dimensional theory.
However, a possible way to naturally overcome the problem of the
scalar ghosts is that the external metric field is invariant under
the action of $G$ in general up to a conformal factor as
\begin{equation}
g_{\mu\nu}(x,\hat{\theta}')=e^{\sigma_g(x,h)}g_{\mu\nu}(x,\hat{\theta}),
\end{equation}
where $\exp\{i\theta'^aT_a\}=\exp\{i\theta^aT_a\}h$, and
$\sigma_g(x,h)$ is a smooth real scalar function on $B^D$. With
the help of the local canonical section taken as
\begin{eqnarray}
s: U&\longrightarrow&\pi^{-1}(U)\nonumber\\
   x&\longmapsto& s(x)=f(x,e),
\end{eqnarray}
we can particularly get an interesting and useful expression for
$g_{\mu\nu}(x,\hat{\theta})$ as follows
\begin{equation}
g_{\mu\nu}(x,\hat{\theta})=e^{\sigma_g(x,\hat{\theta})}g_{\mu\nu}(x).\label{exmetricG}
\end{equation}
In this way, $g_{\mu\nu}(x)=g_{\mu\nu}[s(x)]$ plays the role of
usual $4$-dimensional external gravity field in the effective
$4$-dimensional theory. The warped scalar function
$\sigma_g(x,\hat{\theta})$ determines the conformal diffeomorphism
invariance of the external metric field under the Lie group $G$.
This warped function must clearly satisfy the equation Eq.
(\ref{Gsymm}) but written in the case of the conformal
diffeomorphism invariance. In the low energy limit, this warped
factor will create a vacuum energy density, or cosmological
constant contribution, which is obtained by integrating the last
two terms in Eq. (\ref{E4Dac}) over the vacuum volume of the
internal spaces. Such density is clearly very large because of the
fact that it is proportion to $\Lambda^2$, thus need to be exactly
cancelled to other contributions, for example, the minimum energy
of the moduli stabilization potential
$\widetilde{V}(\widetilde{\phi}_i)$. It is also important to
notice that an interesting consequence of this invariance is that
there will not occur the KK partners of the usual graviton in the
effective low energy theory.This means that the important
phenomenological constraints coming from the KK excitations of the
usual graviton \cite{Giudice1999,Han1999,KKgraexchanges,KKgrapro}
would not transparently happen in the present scenario

Next let us specialize in the physical fluctuations of the modulus
fields by plugging the second expansion of (\ref{Gravityexp}) back
into (\ref{modLag}). Then relevant Lagrangian can be obtained as
\begin{equation}
\mathcal{L}_{\widetilde{h}}=\frac{1}{2}\left(\partial_\mu\widetilde{h}^T\partial^\mu\widetilde{h}
-\widetilde{h}^TM_{\widetilde{h}}\widetilde{h}\right)+\mathcal{L}_{\mathrm{self-int}},
\end{equation}
where $\widetilde{h}$ is a ($D-5$)-component column vector defined
as, $\widetilde{h}=(\widetilde{h}_1...\widetilde{h}_{D-5})^T$,
$M_{\widetilde{h}}$ is a mass mixing matrix among them, and
$\partial^\mu=\eta^{\mu\nu}\partial_\nu$. The physical excitations
describing the curvature perturbation of the internal spaces
correspond with the eigenvectors of $M_{\widetilde{h}}$ whose
masses squared, the eigenvalues of this matrix, are order of
$\Lambda^2$. These massive scalar particles are hidden in the
present world due to the compactification scale, $\sim\Lambda$,
assumed to be very high compared the electroweak scale. They would
be often coupled to the effective $4$-dimensional matter fields
with the coupling strength, $\sim E/M_\textrm{Pl}$, in analogy
with the usual graviton doing. Since these couplings are strongly
suppressed in the energy $E$ to be so much smaller than
$M_\textrm{Pl}$.

\section{\label{4Dfer}$4$-dimensional Weyl fermions on bulk}

In the previously higher dimensional extensions, the fundamental
fermions are given as the spinorial representations of the higher
dimensional Lorentz group. A problem arising from these is that
beside M fermions there would have the presence of very many extra
fermions, which may transform the non-triviality under the gauge
group of SM, in the low energy particle spectrum. Some mechanisms
have been proposed to hide these extra fermions such as the
orbifold compactification
\cite{Pomarol1998,Dienes1999,HCChen2000,Georgi2001} or using the
domain wall \cite{Rubakov1983,Akama1982,Hamed2000}. In the
following we expect to show in our framework that the introduction
of the $4$-dimensional Weyl fermions on the bulk $B^D$ at
fundamental level is a natural result of the intrinsical geometry
of the higher dimensional space-time. The dynamics of these
fermions in the bulk will then be provided in detailed
investigation.

\subsection{Vielbein field}
Let us recall that the $4$-dimensional external tangent space at
each point on $B^D$ spanned by $\{\hat{\partial}_\mu\}$ is endowed
with the given metric $g_{\mu\nu}(x,\hat{\theta})$. However, one
may take the vielbein field ${e_\alpha}^\mu(x,\hat{\theta})$ to
change from the basis $\{\hat{\partial}_\mu\}$ to another basis
$\{\hat{e}_\alpha\}$ as
\begin{equation}
\hat{e}_\alpha={e_\alpha}^\mu(x,\hat{\theta})\hat{\partial}_\mu,
\end{equation}
so that in this basis the $4$-dimensional external metric is
Minkowski-flat. This therefore implies that the vierbein field
must satisfy the following relation
\begin{equation}
\eta_{\alpha\beta}={e_\alpha}^\mu(x,\hat{\theta})
{e_\beta}^\nu(x,\hat{\theta})g_{\mu\nu}(x,\hat{\theta}),\label{loLorentframeI}
\end{equation}
which defines the corresponding local $4$-dimensional Minkowski
space-time at each point on $B^D$. The indices $\alpha$,
$\beta$,... will be used to denote the local $4$-dimensional
Lorentz ones. Note that the metric $\eta_{\alpha\beta}$ here would
not be realized as the perturbative background metric, but is
exactly the metric of the local $4$-dimensional Minkowski
space-time, often called the axial metric. In the formulation of
the vierbein field, the $4$-dimensional external gravity would be
described by ${e_\alpha}^\mu$ instead of the metric $g_{\mu\nu}$.
It is important to note that there are many local $4$-dimensional
Lorentz frames that also yield of the same Eq.
(\ref{loLorentframeI}). Consequently, these will lead to the local
Lorentz transformation rotating among these frames as
\begin{eqnarray}
{e_\alpha}^\mu(x,\hat{\theta})
&\rightarrow&{e'_\alpha}^\mu(x,\hat{\theta})={\Lambda(x,\hat{\theta})^\beta}_\alpha
{e_\beta}^\mu(x,\hat{\theta}),\nonumber \\
\eta_{\alpha\beta}&=&{\Lambda(x,\hat{\theta})^\gamma}_\alpha
{\Lambda(x,\hat{\theta})^\sigma}_\beta\eta_{\gamma\sigma},
\end{eqnarray}
where $\Lambda(x,\hat{\theta})$ is obviously an element of the
Lorentz group SO$(3,1)$. The properly physical significance of
this group in the case of the curved space-time is taken as the
symmetry group of the local $4$-dimensional inertial frames.

From Eq. (\ref{exmetricG}), the expression for the vierbein field
can thus be defined as
\begin{equation}
{e_\alpha}^\mu(x,\hat{\theta})=e^{-\frac{\sigma_g(x,\hat{\theta})}{2}}{e_\alpha}^\mu(x).
\end{equation}
Here ${e_\alpha}^\mu(x)$ satisfies the following relation
\begin{equation}
\eta_{\alpha\beta}={e_\alpha}^\mu(x)
{e_\beta}^\nu(x)g_{\mu\nu}(x),\label{loLorentframeII}
\end{equation}
which defines the local $4$-dimensional inertial frame on an
effective $4$-dimensional space-time manifold. This result also
leads to, $\Lambda(x,\hat{\theta})=\Lambda(x)$, meaning that each
element of the local Lorentz group SO$(3,1)$ only depends on the
external coordinates.

\subsection{Dynamics of $4$-dimensional Weyl fermions}
Next let us proceed to discuss how the $4$-dimensional Weyl
spinors occur on $B^D$. With the novel geometrical structure of
the bulk, we may easily see that they are the simplest non-trivial
irreducible representations of the local Lorentz group SO$(3,1)$
just determined above. This results in that the chiral
$4$-dimensional fermions will occur naturally at the fundamental
level in the present framework. Under the rotation of the local
Lorentz frames, they transform as follows
\begin{eqnarray}
\Psi_L\longrightarrow\Psi'_L&=&e^{\frac{\mathrm{i}}{2}\xi^m\sigma_m}\Psi_L, \\
\Psi_R\longrightarrow\Psi'_R&=&e^{\frac{\mathrm{i}}{2}\eta^m\sigma_m}\Psi_R,
\end{eqnarray}
where $\Psi_{L,R}$ are denoted to the left-handed and right-handed
Weyl spinors written in term of the two-component spinor,
respectively. They are both complex objects depending in general
on the bulk coordinates. Both $\xi^m$ and $\eta^m$, with the index
$m$ running from $1$ to $3$, are complex functions in only the
external coordinates. These function are exactly expressed in the
terms of local transformation parameters, a $4\times 4$
antisymmetric real matrix, of the group SO$(3,1)$. The matrices
$\sigma_m$ are the usual Pauli ones. However, it is almost to work
the Weyl spinors obtained from the projection of the Dirac spinor,
$\Psi_D=(\Psi_L\ \ \Psi_R)^T$, by the operators
$P_{L,R}=(1\mp\gamma_5)/2$. This spinor transforms under the local
Lorentz group SO$(3,1)$ as
\begin{equation}
\Psi_D\longrightarrow\Psi'_D=e^{\frac{i}{2}\epsilon^{\alpha\beta}\Sigma_{\alpha\beta}}\Psi_D,
\end{equation}
where
$\Sigma_{\alpha\beta}=\frac{i}{4}[\gamma_\alpha,\gamma_\beta]$ are
the generators of the Lorentz group SO($3$,$1$) determined in
terms of the usual Dirac matrices $\gamma^\alpha$ obeying the
Clifford algebra,
$\{\gamma_\alpha,\gamma_\beta\}=\eta_{\alpha\beta}$. Their
counterparts in the curved space-time given by
$\gamma_\mu={e^\alpha}_\mu\gamma_\alpha$ satisfying the analogous
one, $\{\gamma_\mu,\gamma_\nu\}=g_{\mu\nu}$. The parameters
$\epsilon^{\alpha\beta}$ of the above local transformation are a
$4\times 4$ antisymmetric matrix with elements which are real
functions in the external coordinates. In what follows we will
work the Weyl spinors expressed via the Dirac spinor.

We would now like to find a suitable Lagrangian describing the
propagation of the $4$-dimensional Weyl spinor fields given above
in $B^D$. We first have to note carefully that with respect to the
present case the usual kinetic term of these fields is impossible
to determine the dynamics in all of directions of the bulk $B^D$.
An obstruction is crucially due to the fact that these fields
would behave the same as a spin-$1/2$ field when moving along the
$4$-dimensional external directions but as a scalar field when
moving in the internal directions. The scalar-like behavior is
realized in sense of that the usual spin concept is lost under the
viewpoint of the internal evolution in which instead of this they
carry out the conventionally internal charges inherited from their
non-trivial representations under the group SO$(3,1)$. This means
that the physical manifestation is distinctly different to
correspond with the dynamics taken in the external and internal
spaces. This analysis thus give a better understanding of their
dynamic properties in $B^D$ which will be reasonable to construct
a consistent Lagrangian. Notice that it can also be
straightforward for the more generic case in which fields have an
arbitrary spin. A full Lagrangian thus includes two independently
separating parts which determine the evolution along the external
and internal directions, respectively. The evolution of such a
Weyl spinor field with specifically given chirality along the
$4$-dimensional external directions are characterized by the usual
Lagrangian
\begin{equation}
\mathcal{L}^\Psi_{4D}=\bar{\Psi}i\gamma^\alpha {e_\alpha}^\mu
D_\mu\Psi,\label{exPsiLa}
\end{equation}
where $\Psi$ is denoted to the (right-handed) left-handed Weyl
spinor field with the dimension $[\Psi]=\frac{D-1}{2}$. In
particular, it may be dimensionally renormalized in the natural
way as
\begin{equation}
\Psi_{\textrm{nor}}=\frac{\Psi}{\Lambda^\frac{D-4}{2}},
\end{equation}
where the factor $1/\Lambda^\frac{D-4}{2}$ is taken from the bulk
volume element. The redefinition above leads to
$[\Psi_{\textrm{nor}}]=\frac{3}{2}$ which is equal to the
dimension of the fermion in the $4$-dimensional
space-time.\footnote{ Other fundamental fields living on $B^D$ can
also be taken similarly by such way to have the same dimension as
the corresponding field in the $4$-dimensional theory. By this, it
is interesting to observe that the dimension of coupling constants
in the higher dimensional theory, e.g. as that for the gauge,
Yukawa interactions or the self-couplings of the scalar fields,
would be respectively same as in the usual $4$-dimensional
theory.} It is important to note that this redefined field
$\Psi_{\textrm{nor}}$ is not treated as the infraredly effective
field, but it is still the fundamental field. This is because the
redefinition is formally independent on a cutoff scale at which
the corresponding low energy description will break down. In the
following discussion we will be interesting in the renormalized
fields as above in which ``$\textrm{nor}$'' will be dropped. On
the other hand, the factor $1/\Lambda^{D-4}$ in the original
matter action is completely eliminated through the redefinition of
the field. The covariant derivatives $D_\mu$ acting $\Psi$ are
given by
\begin{equation}
\hat{D}_\mu=\hat{\partial}_\mu+\frac{i}{2}\omega^{\alpha\beta}_\mu\Sigma_{\alpha\beta},
\label{spcd}
\end{equation}
where $\omega^{\alpha\beta}_\mu$ is the torsion-free spin
connection which is expressed in terms of the vierbein field as
\begin{equation}
\omega^{\alpha\beta}_\mu={e^\alpha}_\nu\left(\hat{\partial}_\mu
e^{\beta\nu}+e^{\beta\lambda}\Gamma^\nu_{\mu\lambda}\right),
\end{equation}
with ${e^\alpha}_\nu$ to be inverse of the vierbein field
${e_\alpha}^\nu$, and
$e^{\beta\nu}=\eta^{\beta\gamma}{e_\gamma}^\nu$. It is important
to notice that the $4$-dimensional Weyl spinor fields moving along
the $4$-dimensional external directions with the specific
chirality are massless due to their mass term vanishing. Next, let
us attempt to construct an internal Lagrangian for $\Psi$. As just
discussed above, in order to do this we must first find out the
bilinear terms of $\Psi$ to be a Lorentz scalar which satisfying
only consist of, $\bar{\Psi}^c\Psi$. Here $\Psi^c=C\Psi^{*}$ is
the charge conjugate field of $\Psi$, with $*$ denoted to the
complex conjugate and $C$ to be a $4\times 4$ matrix satisfying
the following conditions, $C^\dag C=1$ and $C^\dag\gamma^\mu
C=-\gamma^{\mu*}$. It should be noted that an another term mixing
between the left-handed and right-handed Weyl spinors is not
included by the property of the chirality in which they transform
differently under the local gauge group. This term is thus
forbidden obviously, and only generated via the coupling to the
Higgs field after this field gets non-zero VEV. From the above
argument, we can find out the following expectant Lagrangian
\begin{equation}
\mathcal{L}^\Psi_{G}=\frac{1}{2\Lambda}\left(-\hat{\partial}_i\bar{\Psi}^c\hat{\partial}^i\Psi-M^2_\Psi\bar{\Psi}^c\Psi\right)+\textrm{H.c.},\label{inLagWeylspI}
\end{equation}
where real number $M^2_\Psi$ appearing in the above Lagrangian
plays the role of mass squared characterizing to the internal
motion. It thus is naturally taken to the same order as the scale
$\Lambda$. We can see that this Lagrangian being analogous to that
of the scalar field is quadratic in the internal derivatives
$\hat{\partial_i}$. At first important sight, all of the terms in
the Lagrangian (\ref{inLagWeylspI}) vanish identically if $\Psi$
is realized classically as the ordinary commuting field, or
$c$-number. This is due to the presence of the totally
antisymmetric tensor $\epsilon_{ab}$ (with $a,b=1,2$ and
$\epsilon_{12}=1$) in components of the bilinear combinations of
$\Psi$. This means that the internal dynamics of the
$4$-dimensional Weyl spinor fields does not really make sense in
the classical theory, but is of the quantum-mechanical concept in
which they should be anticommuting fields. At second important
one, $\Psi$ in the Lagrangian (\ref{inLagWeylspI}) is not able to
play the role of the physical field under the internal point of
view. This is because each term of (\ref{inLagWeylspI}) is not
hermitian leading to that the corresponding Hamiltonian is
unbounded from below. A combination of $\Psi$ and $\Psi^c$,
however, defined as
\begin{equation}
\Psi_M=\Psi+\Psi^c,
\end{equation}
provides the well-defined physical field. Therefore,
$\mathcal{L}^\Psi_{G}$ is physically rewritten as
\begin{equation}
\mathcal{L}^\Psi_{G}=\frac{1}{2\Lambda}\left(-\hat{\partial}_i\Psi^T_MC\hat{\partial}^i\Psi_M-M^2_\Psi\Psi^T_MC\Psi_M\right).\label{inLagWeylspII}
\end{equation}
The field $\Psi_M$ is known as the Majorana spinor field (the real
Dirac spinor) having the degrees of freedom to be equal to those
of $\Psi$. It satisfies the reality condition, $\Psi_M=\Psi^c_M$,
being invariant under the Lorentz transformation.\footnote{Notice
that, the Dirac matrices $\gamma^\mu$ transform under a group of
the unitary matrices as, $\gamma^\mu\longrightarrow U\gamma^\mu
U^\dag$, corresponding to the rotation of the basis of the
Clifford algebra. Using this transformation, one can look at a
novel basis in which $C=1$ and all $\gamma^\mu$ are pure imaginary
known as the \emph{Majorana basis}. Hence the reality condition of
the field $\Psi$ is now simple as, $\Psi^c=\Psi^*=\Psi$, which is
precisely same as that for a scalar field. The corresponding
internal Lagrangian is thus of that of the real scalar field.} The
Lagrangian (\ref{inLagWeylspII}) describes the physical progress
of a scalar-like neutral Grassman field in the internal spaces in
which it carries out the globally SO$(3,1)$ internal symmetry in
the usual sense. The structure of the KK tower for $\Psi_M$ is
determined through the operator, $(\square_G-M^2_\Psi)/\Lambda$.
Since the lowest level has non-zero energy of the value
$M^2_\Psi/\Lambda$.

In the fact, there exists a significant coupling between the
modulus fields to $\Psi$ which is given by the non-trivial
invariant terms in Eq. (\ref{modPotI}) coupled to
$\bar{\Psi}^c\Psi$. At leading order, this coupling form is taken
by
\begin{equation}
\sum^{D-5}_{i}\lambda_{\widetilde{\phi}_i\Psi}\widetilde{\phi}_i\bar{\Psi}^c\Psi+\textrm{H.c},\label{Psimodint}
\end{equation}
where $\lambda_{\widetilde{\phi}_i\Psi}$ are dimensionless
coupling constants. Notice that, the higher-order terms in
$\widetilde{\phi}_i$ in the coupling above are suppressed by the
positive powers of $\overline{M}_P$ while the zero-order term
leading to the mass term for $\Psi$ must obviously be zero.

It is worth stressing here one remarkable property of the internal
Lagrangian given above. It may be easily verified that the
Lagrangian (\ref{inLagWeylspI}) is only allowed with respect the
fields which are neutral under all the local gauge charges, such
as the right-handed neutrinos occurring in extensions of SM. In
other words, this is not of the case of the fields of carrying out
such charges, for example, the fermions of SM, due to that their
Lagrangian would break the local gauge symmetry explicitly. If a
$4$-dimensional Weyl spinor field $\Psi$ (for what is analyzed in
the following we will still use $\Psi$ to refer the field under
consideration) has the exactly conserved quantum charges, then its
internal Lagrangian is precisely forbidden. Consequently, this
results in that such field is by itself active (conformal)
diffeomorphism invariance under the action of $G$. The values of
$\Psi$ on the same internal space are thus related together. So by
analogy with what has been performed for the $4$-dimensional
external metric field, we can obtain an useful expression for
$\Psi$ as
\begin{equation}
\Psi(x,\hat{\theta})=D(x,\hat{\theta})\psi(x).\label{GpsiA}
\end{equation}
Here $D(x,\hat{\theta})$ which is a scalar
function\footnote{$D(x,\hat{\theta})$ is in general a matrix in
the representation space of $\Psi$. However, in order for Eq.
(\ref{GpsiA}) well-defined, it must indeed be independent of the
the above local Lorentz transformation group as well as the gauge
transformations arising in the realistic model. Since
$D(x,\hat{\theta})$ must commutate with all the generators of the
groups which the field $\Psi$ transforms the non-triviality. This
means that $D(x,\hat{\theta})$ is necessary a matrix to be
proportional to the identity matrix.} presents the \emph{intrinsic
diffeomorphism invariance} of $\Psi$ under the action of $G$ and
obviously satisfies Eq. (\ref{Gsymm}). In low energy limit,
$D(x,\hat{\theta})$ is approximately to only depend on the
internal coordinates. The field $\psi(x)$ encodes the
$4$-dimensional behavior of the fundamental field $\Psi$.
Consequently, we assume that $\Psi$ carries additionally out the
usual local gauge symmetry $G'$, then each element of the gauge
group $G'$ does not depend on the internal coordinates. The gauge
fields corresponding with the gauge group $G'$ have thus only the
$4$-dimensional external components denoted by $B_\mu$ which are
by themselves active diffeomorphism invariance under the action of
$G$ in general up to a conformal factor. This is due to that their
internal Lagrangian built uniquely as
\begin{equation}
\sim
g^{\mu\nu}\mathrm{Tr}[\hat{\partial}_iB_\mu\hat{\partial}^iB_\nu],
\end{equation}
is forbidden by violating the gauge invariance.

\subsection{Application to explain the smallness of the neutrino masses}

Let us provide a phenomenologically significant consequence of the
dynamical description for the $4$-dimensional Weyl spinor fields
above. It is well established by the experimental observations
that the masses of the neutrinos are so much smaller than those of
the charged leptons and quarks in SM. We now desire to provide a
possibility to realize the natural origin of this problem. We
consider the minimal extension of SM given by including three
right-handed neutrinos $\nu_R$ associated with three lepton
families. They are all obviously singlet under the gauge group of
the SM. As seen above, all of the charged leptons, the quarks and
the gauge bosons of the SM are diffeomorphism invariance by
themselves under the action of the Lie group $G$. Only the
right-handed neutrinos and the SM Higgs doublet are to have the
non-trivial dynamics in the internal spaces. The electroweak
symmetry is spontaneously broken by the VEV of this scalar field
whose breaking scale is possible to be much lower than the
compactificated one. Hence a mass Lagrangian for the neutrinos in
the flavor basis is taken in the effective low energy description
as
\begin{equation}
\mathcal{L}^{\textrm{mass}}_{\nu}=-\sum_{n}\left(m_{n}\bar{\nu}_L\nu_{nR}+\frac{1}{2}M_n\bar{\nu}^c_{nR}\nu_{nR}\right)+\textrm{H.c.},\label{NML}
\end{equation}
where $\nu_{nR}$ is the $n$th KK mode of the right-handed
neutrinos, and all of the fermions have been normalized
canonically. Notice that the flavor indices have been suppressed
for convenience. The Dirac masses are defined by
\begin{equation}
m_n=\kappa_nv_{\textrm{w}},\label{DNmasses}
\end{equation}
where $v_\textrm{w}$ is the electroweak symmetry breaking scale
corresponding to the VEV of the SM Higgs doublet. The
dimensionless coefficient $\kappa_n$ contains the information
about the corresponding bulk Yukawa coupling constant, the
functions determining the diffeomorphism invariance of the
left-handed neutrinos under $G$ and the profile wave function of
the $n$th KK mode of $\nu_R$. Such constants are effectively
matched as $4$-dimensional Yukawa coupling constants of the
neutrinos. The Majorana masses occurring in a natural way in the
$4$-dimensional effective theory are taken as
\begin{equation}
M_n=\frac{M^2_R}{\Lambda}+\frac{\lambda^2_n}{\Lambda}+\sum^{D-5}_{i}\lambda_{\widetilde{\phi}_i\nu}\langle\widetilde{\phi}_i\rangle.\label{MNmasses}
\end{equation}
The mass $M^2_R/\Lambda$ comes from the Lagrangian describing the
internal dynamics for the right-handed neutrinos. The second term
is mass of the $n$th KK right-handed neutrinos. The contribution
by the last term is produced by the coupling between the
right-handed neutrinos to the modulus fields whose form is given
by (\ref{Psimodint}). Therefore, the active neutrinos acquire the
masses through the $5$-dimensional effective operator
\cite{Weinberg1979} corresponding to the KK modes of the
right-handed neutrinos integrated out as
\begin{equation}
m^{\mathrm{ac}}_\nu=m_DM^{-1}m^T_D.\label{acneumassI}
\end{equation}
Here $m_D$ and $M$ are ($1\times\infty$) Dirac and
($\infty\times\infty$) Majorana mass matrices, respectively,
constructed from the above mass expressions. We can approximately
ignore the mixing between the left-handed neutrinos and the KK
excitations of the right-handed neutrinos, thus they decouple to
the mass spectrum of the active neutrinos given at the lowest
order
\begin{equation}
m_\nu\cong\frac{m^2_{0D}}{M_0}.
\end{equation}
Eventually, the smallness of the neutrino masses at the sub-eV
scale to be consistent with the neutrino oscillation data is
realized via the type I seesaw mechanism which requires the scale
$M_0$ $\sim 10^{6}$ GeV if the Dirac masses $m_{0D}$ of the
neutrinos are of order of the electron mass. Therefore, we can say
that the small mass appearance of the observed neutrinos is just
of consequences of the higher dimensional space-time geometry.

\section{\label{gaugecharges}Fields with internal gauge charges}

In this section, we will proceed to be mainly interesting in
fields on $B^D$ carrying out remarkable representations of the
symmetry group $G$ of the internal spaces. These fields will
correspond to the conventional source for the gauge fields
$A^a_\mu$. Let us consider a field $\Phi(x,\hat{\theta})$ which is
scalar with respect to both the local external coordinate
transformation group $\textrm{GL}(4,\mathbb{R})$ and the local
Lorentz group SO$(3,1)$, but carries out a $d$-dimensional
representation of the Lie group $G$. This multiple is expressed in
term of a column vector with $d$ scalar fields. Under the local
coordinate transformations (\ref{gct}), the field $\Phi$ rotates
as follows
\begin{equation}
\Phi(x,\hat{\theta})\rightarrow\Phi'(x',\hat{\theta}')=U\Phi(x,\hat{\theta}),\label{PhiTr}
\end{equation}
where, $U=D[h(x)]=\textrm{exp}\{i\alpha^a(x)M_a\}$, is a $d\times
d$ matrix corresponding to a representation of the element
$h(x)\in G$ given in (\ref{gct}) at which $M_a$ are also $d\times
d$ matrices associated with the representation of the generators
$T_a\in\mathfrak{g}$. We can make an observation that the
transformation (\ref{PhiTr}) is nearly analogy to that in the
usual gauge theory, but here it has a deep connection with the
internal structure of the higher dimensional space-time. Thus, the
presence of such fields in Nature would be specially useful for
better intuitive understanding about the internal symmetric
aspects of the higher dimensional space-time. Moreover, they would
play the crucial role as the important source for the physical
kinetics of $A^a_\mu $ to generate the topological twisting of the
bundle $B^D$. A geometrical interpretation of such field is taken
as a local section on the associated vector bundle with the base
space $B^D$ rather than $M^4$ as in the original gauge theory.

The physical progress of the field $\Phi$ in the bulk are
determined by the following Lagrangian
\begin{equation}
\mathcal{L}_\Phi=(D^\mu\Phi)^{\dag}(D_\mu\Phi_i)-(\hat{\partial}^i\Phi)^{\dag}(\hat{\partial}_i\Phi)
-V(\Phi),
\end{equation}
where the external covariant derivatives $D_\mu$ are defined in a
natural way by the gauge fields $A^a_\mu$ on $B^D$ as
\begin{equation}
D_\mu=\hat{\partial}_\mu+ig_\textrm{i}A^a_\mu M_a.
\end{equation}
The above given Lagrangian to be invariant under (\ref{PhiTr})
gives rise the gauge transformation rule as
\begin{equation}
A'^a_\mu(x')M_a=\frac{\partial x^\nu}{\partial
x'^\mu}\left[UA^a_\nu(x)M_aU^{-1}
+\frac{1}{ig_\textrm{i}}U\partial_\nu
U^{-1}\right].\label{gautrnII}
\end{equation}
This is to correspond with the transformation (\ref{gautrnI}), but
written in the form of the given representation. The infinitesimal
form of (\ref{gautrnII}) can be obtained with the same result as
in Eq. (\ref{infgautr}).

The scalar potential $V(\Phi)$ is introduced in the following form
\begin{equation}
V(\Phi)=\mu^2\Phi^\dag\Phi+\lambda(\Phi^\dag\Phi)^2.\label{Phipo}
\end{equation}
The mass squared $\mu^2$, as analyzed above, involves two
independent masses squared, meaning that
$\mu^2=\mu^2_{\mathrm{ex}}+\mu^2_{\mathrm{in}}$, in which
$\mu_{\mathrm{ex}}$ and $\mu_{\mathrm{in}}$ are the characteristic
masses for the dynamics in the external and internal spaces,
respectively. The quartic-order coupling constant $\lambda$ is
real and dimensionless. We have also assumed that trilinear
couplings of $\Phi$ violates the gauge invariance (\ref{PhiTr})
explicitly, since these are absent in the above potential. In
addition, the field $\Phi$ of course couples to the modulus fields
(in such a way to be similar as the field $\Psi$ discussed in the
previous section coupled) and the SM Higgs doublet $H$ taken as
\begin{equation}
\left(\sum^{D-5}_{i}\lambda_{i}\widetilde{\phi}_i+\sum^{D-5}_{i,j}\lambda_{ij}\widetilde{\phi}_i\widetilde{\phi}_j\right)\Phi^\dag\Phi,\
\ \lambda_{\Phi H}\Phi^\dag\Phi H^\dag H,\label{Phi-mod-Hcoup}
\end{equation}
where the coupling constants $\lambda_{i}$ have the mass dimension
whereas $\lambda_{ij}$ and $\lambda_{\Phi H}$ are dimensionless
ones.

Due to the absence of the Yukawa couplings between $\Phi$ with the
fermions given in the preceding section which are not allowed by
the transformation (\ref{PhiTr}),\footnote{The quantum anomaly
often caused by the fermions transforming non-triviality under the
local gauge symmetry is automatically cancelled in this
framework.} so the total Lagrangian possesses an accidentally
exact $Z_2$-discrete symmetry
\begin{equation}
\Phi\longrightarrow -\Phi.\label{Z2symm}
\end{equation}
Without loss of generality, other fields can be taken to transform
the trivial way (or being even) under this symmetry. Clearly, this
is to come of both the symmetries of the higher dimensional
space-time and the renormalizable coupling terms.

In analogy with the standard procedure to construct the kinetic
term for the gauge fields in the conventional gauge theory, we
need first to define the covariant field strength tensor. The
easiest way to derive this tensor is through the commutator of
covariant derivatives
\begin{equation}
\hat{F}_{\mu\nu}=\frac{1}{g_{\textrm{i}}}[D_\mu,D_\nu]=-\frac{1}{\Lambda}F^i_{\mu\nu}\hat{\partial}_i+iF^a_{\mu\nu}M_a.
\end{equation}
The Lagrangian for $A^a_\mu$ can then be given as
\begin{eqnarray}
\mathcal{L}_A&=&\frac{1}{2}\left(\frac{M_*}{\Lambda}\right)^{D-4}\textbf{Tr}[\hat{F}_{\mu\nu}\hat{F}^{\mu\nu}]\nonumber
\\
&=&\left(\frac{M_*}{\Lambda}\right)^{D-4}\left(-\frac{g^2_{\textrm{i}}M^2_*}{4\Lambda^2}\sum_{i}e^{\widetilde{\phi}_i/\overline{M}_P}F^i_{\mu\nu}F^{i\mu\nu}-\frac{1}{4}F^a_{\mu\nu}F^{a\mu\nu}\right),\label{ALag}
\end{eqnarray}
where $\textbf{Tr}$ refers a symmetric inner product determined by
\begin{equation}
\textbf{Tr}[\hat{\partial}_i\hat{\partial}_j]=-\frac{g^2_{\textrm{i}}M^2_*}{2}\delta_{ij}e^{\widetilde{\phi}_i/\overline{M}_P},\
\ \textbf{Tr}[M_aM_b]=\frac{\delta_{ab}}{2}, \  \
\textbf{Tr}[\hat{\partial}_i M_a]=0.
\end{equation}
Comparing with (\ref{YMaction}) and the last term in (\ref{EHa}),
we can easily see that the last expression of (\ref{ALag}) agrees
with the dynamical Lagrangian for the gauge fields $A^a_\mu$.

Before proceeding to come the last section, we would like to
suggest an extremely natural candidate for the dark matter (DM)
coming from the scalar multiple $\Phi$ above if it is neutral
under the gauge symmetry of SM. We consider the case which the $G$
gauge sector is at broken phase to be consistent with the
observable fact whose scale would be higher than the electroweak
scale. To maintain the multiple $\Phi$ which plays the role in the
dark matter, in addition to this we should introduce a different
multiple $\Phi'$ to break spontaneously the gauge symmetry and
generate masses for the gauge bosons $A^a_\mu$. Since the massive
gauge bosons $A^a_\mu$ are hidden in the present observed world.
For example, with $G$ to be the Lie group $\mathrm{SU}(2)$ as we
already discussed previously, one can easily check a possibility
that $\Phi$ and $\Phi'$ are taken as the double and triplet of
this group, respectively. This means that the vacuum would be
chosen spontaneously as, $\langle\Phi\rangle=0$ (thus $\Phi$ would
be realized as the inert multiple) and $\langle\Phi'\rangle\neq0$,
which is assumed to exist. It is of the vacuum configuration
described by the following surface
\begin{equation}
F_{\textrm{vac}}(\langle\Phi\rangle,\langle\Phi'\rangle)=0.\label{Evacon}
\end{equation}
This associates with much infinitely degenerate ground states
defined classically via the vanishing of the first-order
derivatives of the potential $V(\Phi,\Phi')$. Note that due to the
potential and its space of minima having the same symmetry since
this minimal surface must of course have the isometric group $G$.
As a consequence, the $Z_2$ symmetry given in (\ref{Z2symm})
remains to be conserved by $\Phi$ which guarantees the stability
with respect the lightest particle of the old charge under this
parity. There has in general the mass splitting in $\Phi$ by
coupling to $\Phi'$. Hence the lightest component in the zero-mode
of the multiple $\Phi$ is responsible to the dark matter. In this
case, the SM Higgs field and the massive gauge bosons $A^a_\mu$
provides predominantly the portal which links the visible sector
(the SM particles) and the dark sector whose origin is to come
from the extension of the space-time structure. Notice that, these
two sectors can also communicate together via higher dimension
operators. Since the DM pair annihilation into the light SM
particles and the DM scattering on nuclei are transmitted via both
the Higgs field and $A^a_\mu$. It is important that if mass of the
DM candidate is smaller than the half of the mass of Higgs boson
determined by ATLAS and CMS collaborations \cite{ATLAS,CMS}, it
would contribute in the invisible decay of the Higgs bosons with
the current experimental data at LHC given in \cite{Belanger2013}
leading to constrains on its mass and couplings. It is interesting
to see that the stability of the dark matter is fully guaranteed
by the intrinsically dynamical symmetry of theory rather than the
symmetry imposed by the hand. To show the above given analysis
more clearly, a detail study of dark matter phenomenology and
constraints will be taken in \cite{Darkmatter}.

\section{\label{conclus}Conclusions and Comments}

This work has presented the importantly physical implications
coming from a remarkably geometrical form of the $D$-dimensional
space-time manifold $B^D$ which is assumed to occur at high energy
region. It is a principal bundle constructed by attaching a smooth
copy of the ($D-4$)-dimensional Lie group $G$, considered in this
paper as $\mathrm{SU}(N)$ group, at each point of a
$4$-dimensional pseudo-Riemannian manifold with the Lorentz
metric. On the other hand, under dynamics motivated by the sources
$B^D$ is always foliated by ($D-4$)-dimensional submanifolds
deforming smoothly of $G$. These submanifolds are thought of as
the internal spaces (or fibres) while the submanifolds being
transversal to the fibres are realized as the external spaces
which describe the usual $4$-dimensional world. The geometrical
dynamics of $B^D$ is thus realized to associate with the high
energy gravity. The corresponding fundamental degrees of freedom
in the most minimal scheme consist of the gauge fields $A^a_\mu$,
the $4$-dimensional external metric field $g_{\mu\nu}$ and the
modulus fields $\phi_i$, $i=1,..,D-5$, which set dynamically the
volume of the internal spaces. The presence of the gauge fields
$A^a_\mu$ is to point precisely out the local directions of the
external spaces which depend on the topological non-triviality of
the bundle $B^D$. Consequently, an arbitrary particle moving along
the external directions would be possible to interact with
$A^a_\mu$. The corresponding gauge charges are generated
dynamically by the evolution in the internal spaces. In this way,
our framework is an extension of the gauge interaction in
connecting closely to the structure of the space-time. All these
fields are unified in the same non-trivial geometrical framework
of the higher dimensional space-time. Remarkably, the potential of
the moduli stabilization is constructed in a natural way that
gives rise the mechanism to fix dynamically the size of the
internal spaces. This is well done due to that the Lie group $G$
acts freely on $B^D$ and transitively on each internal space. It
has been explicitly shown, for instance, the Lie group
$\mathrm{SU}(2)$. By the above moduli stabilization potential, the
modulus excitations have also heavy masses. The gauge fields
$A^a_\mu$ get masses to become massive via the Higgs mechanism
corresponding with the gauge symmetry group $G$ spontaneously
broken. In this way, the mechanism to hidden the effects of the
extra dimensions in the experimental seeking on accelerators is
very natural. Therefore, we are completely not worry the existence
of massless particles in the effective low energy theory which
would lead to the unwanted violations of the equivalence principle
because of contributing to Newton's law. In summary, at the large
distances, the gravity is well described by general relativity at
which gravitational effects originating from extra dimensions are
strongly suppressed. However, as the energy increasing highly, the
usual $4$-dimensional classical gravity will get modification from
such high energy effects.

We have further studied the dynamics of the $4$-dimensional Weyl
spinor fields laying $B^D$ which are the simplest non-trivial
representations of the Lorentz group $\mathrm{SO}(3,1)$. This
group is determined as the local symmetry group of the local
$4$-dimensional inertial frames corresponding with the
Minkowski-flat form of the tangent spaces of the external spaces.
The natural existence of these fields on $B^D$ would thus yield
the very favorable light to overcome the chiral fermion problem
occurring almost in the $4$-dimensional effective theory matched
from the higher dimensional theory. An interesting result
inherited from this description is that their physical behaviour
is manifested in a distinctly different way under the dynamics in
the $4$-dimensional external and internal spaces. Strictly
speaking, observers will see them to behave as the spin-$1/2$
particles under the usual $4$-dimensional external point of view
but as the scalar-like particles with respect to the internal
point of view (i.e. their spin would be lost, and instead of this
they carry out the conventional internal symmetry of the Lorentz
group $\mathrm{SO}(3,1)$). Based on this analysis, the external
Lagrangian for these fields has been found with the usual Dirac
one but written in the curved space-time. Whereas the internal
Lagrangian has the form in analogy to that of the real scalar
field, but only making quantum sense without the classical
equivalence. There is an important consequence related to this
construction at which the internal dynamic Lagrangian is forbidden
with respect to the $4$-dimensional Weyl spinor fields of carrying
out the usually local gauge charges. Such fields are thus imposed
by themselves the active (conformal) diffeomorphism invariance
under the action of the Lie group $G$. In particular, another
interesting result derived provides a possible natural
interpretation for the tiny mass origin of the neutrinos observed
in the present experiments via the type I see-saw mechanism.

The inclusion of the fields transforming a non-trivial way under
the local symmetry group of the internal spaces is investigated in
detail. They are constructed in a similar way as in the context of
the usual gauge theory. A geometrical description for these fields
is given as the local sections of the associated vector bundle on
the $B^D$ rather than on the base space of $B^D$. The
$4$-dimensional Weyl fermions discussed above are obviously
neutral under this group. We also suggest a natural candidate for
the dark matter coming from these novel fields to be neutral under
the gauge group of the SM. An accidentally exact $Z_2$-discrete
symmetry conserved in the broken phase of the $G$ gauge symmetry
protects its stability.

Beyond those, a further issue for next work is to consider the
supersymmetric extension in which the bulk space-time is included
additionally the anticommuting extra dimensions
\cite{NamSUSY2014}. This will be specially interesting by the fact
that $B^D$ contains both $4$D Poincar\'{e} and gauge symmetries.
Therefore, the rules of building the suppersymmetry in the
previously higher dimensional space-time is clearly impossible to
apply in such case.

Finally, before ending this discussion we wish to take the
following useful comment. Although we have been only interesting
in the internal spaces which are equivalent smoothly to the
compact connected Lie group, the above given analysis will be able
to be extended to some the manifolds which do not carry out the
structure of the Lie group. As well known, quite many manifolds
are constructed as the coset space $G/H$ that is the quotient of
Lie group $G$ and its Lie subgroup $H$. If we identify physically
at the points on the bulk $B^D$ which belong the same orbit under
the action of $H$, then this result in the internal spaces which
are the smooth copies of $G/H$. Note that, because the action of
$H$ is free so the resulting internal spaces are also smooth
manifolds unless the above identification is taken by rather the
discrete subgroups than the continuous ones leading to the
internal spaces with singularities, or well-known as orbifolds.
These have been used in the scenarios of particle physics with
extra dimensions to get the low energy realistic particle spectrum
as well as the gauge symmetry or suppersymmetry breaking.
\section*{Acknowledgments}
I would like to thank Dr. Le Tho Hue at Institute of Physics,
Vietnam Academy of Science and Technology for useful and
stimulating discussions.

\appendix*

\section{Non-holonomic functions, Christoffel symbols
and bulk scalar curvature}

The frame fields on the bulk are given by,
$\{E_M(X)\}=(\{\hat{\partial}_\mu\},\{\hat{\partial}_i\})$. The
non-holonomic functions $C^P_{MN}$ are then determined through the
Lie bracket of any two frame fields as
\begin{equation}
[E_M(X),E_N(X)]=C^P_{MN}E_P(X),
\end{equation}
whose explicit expression is defined as follows
\begin{eqnarray}
C^\lambda_{\mu\nu}&=&0,\hspace*{0.5cm} C^i_{\mu\nu}=-\frac{g_\textrm{i}}{\Lambda}F^i_{\mu\nu},\label{C1} \\
C^\nu_{\mu i}=-C^\nu_{i\mu}&=&0,\hspace*{0.5cm}C^j_{\mu i}=-C^j_{i\mu}=0,\label{C2}\\
C^k_{ij}&=&\overline{f}^k_{ij}.\label{C3}
\end{eqnarray}

With the bulk metric given generally in Eq. (\ref{bmt}), the
explicit expression for the Christoffel symbols as
\begin{eqnarray}
\Gamma^\rho_{\mu\nu}&=&\frac{g^{\rho\lambda}}{2}\left(\hat{\partial}_\mu
g_{\nu\lambda}+\hat{\partial}_\nu
g_{\mu\lambda}-\hat{\partial}_\lambda g_{\mu\nu}\right),\\
\Gamma^i_{\mu\nu}&=&\frac{1}{2}\left(\gamma^{ij}\hat{\partial}_jg_{\mu\nu}-
\frac{g_\textrm{i}}{\Lambda}F^i_{\mu\nu}\right), \\
\Gamma^\nu_{\mu i}&=&\Gamma^\nu_{i\mu
}=\frac{g^{\nu\lambda}}{2}\left(\hat{\partial}_ig_{\mu\lambda}-
\frac{g_\textrm{i}}{\Lambda}\gamma_{ij}F^j_{\mu\lambda}\right),\\
\Gamma^j_{\mu
i}&=&\Gamma^j_{i\mu}=\frac{1}{2}\gamma^{jk}\hat{\partial}_\mu\gamma_{ik},\\
\Gamma^\mu_{ij}&=&\frac{1}{2}g^{\mu\nu}\hat{\partial}_\nu\gamma_{ij},\\
\Gamma^k_{ij}&=&\frac{1}{2}\left[\gamma^{kl}\left(\overline{f}^m_{li}\gamma_{mj}+\overline{f}^m_{lj}\gamma_{mi}\right)+\overline{f}^k_{ij}\right].
\end{eqnarray}

The scalar curvature of the bulk is defined by
\begin{eqnarray}
R&=&\hat{R}+\frac{1}{4}\hat{\partial}_ig^{\mu\nu}\hat{\partial}^ig_{\mu\nu}-\frac{1}{4}\left(g^{\mu\nu}\hat{\partial}_ig_{\mu\nu}\right)\left(g^{\rho\lambda}\hat{\partial}^ig_{\rho\lambda}\right)
-\frac{1}{4}\hat{\partial}_\mu\gamma^{ij}\hat{\partial}^\mu\gamma_{ij}
\nonumber \\
&&+\frac{1}{4}\left(\gamma^{ij}\hat{\partial}_\mu\gamma_{ij}\right)\left(\gamma^{kl}\hat{\partial}^\mu\gamma_{kl}\right)
-V(\gamma_{ij})-\frac{g^2_\textrm{i}}{4\Lambda^2}\gamma_{ij}F^i_{\mu\nu}F^{j\mu\nu}+\sum^{4}_{i=1}\nabla_M
X^M_i,
\end{eqnarray}
where $\hat{R}$ is the 4-dimensional standard scalar curvature
given in Eq. (\ref{4Dscur}), the potential $V(\gamma_{ij})$ for
the internal metric corresponding to the scalar curvature of the
internal spaces is given as
\begin{equation}
V(\gamma_{ij})=\Lambda^2\left(-\frac{1}{2}f^k_{il}f^l_{kj}\gamma^{ij}
+\frac{1}{4}f^p_{ik}f^q_{jl}\gamma^{ij}\gamma^{kl}\gamma_{pq}\right),
\end{equation}
and
\begin{eqnarray}
X^M_1&=&\left(-\frac{g^{\mu\nu}\gamma^{ij}\hat{\partial}_\nu\gamma_{ij}}{2}, 0\right), \  \ X^M_2=\left(0, \Lambda\frac{\gamma^{ij}g^{\mu\nu}\partial_jg_{\mu\nu}}{2}\right), \nonumber \\
X^M_3&=&\left(\frac{g^{\mu\nu}\gamma^{ij}\hat{\partial}_\nu\gamma_{ij}}{2},0\right),\  \ X^M_4=\left(0,-\Lambda\frac{\gamma^{ij}g^{\mu\nu}\partial_jg_{\mu\nu}}{2}\right). \nonumber \\
\end{eqnarray}

\end{document}